\documentclass[usegraphicx,useAMS,usenatbib]{mn2e}
\usepackage[english]{babel}
\usepackage[utf8x]{inputenc}
\usepackage{amsmath}
\usepackage{amssymb}
\usepackage{latexsym}
\usepackage{graphicx}
\newcommand{\fedd}{$f_{\rm Edd}$}

\newcommand{\ho}{$H_0$}

\newcommand{\kmsm}{km\,s$^{-1}$\,Mpc$^{-1}$}
\newcommand{\mbh}{$M_{BH}$}

\newcommand{\apj}{Astrophys.~J.}
\newcommand{\apjl}{Astrophys. J. Lett.}
\newcommand{\aap}{Astron. Astrophys.}
\newcommand{\mnras}{Mon. Not. R. Astron. Soc.}
\newcommand{\pasj}{Publ. Astron. Soc. Jpn.}

\begin{document}
\title{Modeling quasar accretion disc temperature profiles}
\author[Hall et al.]{P. B. Hall$^{1}$\thanks{E-mail: phall@yorku.ca},
E. S. Noordeh$^1$, L. S. Chajet$^1$, E. Weiss$^1$,
C. J. Nixon$^{2,3}$\\
$^1$Department of Physics and Astronomy, York University, Toronto, ON M3J 1P3, Canada\\
$^2$JILA, University of Colorado \& NIST, Boulder, CO 80309-0440, USA\\
$^3$Einstein Fellow\\
}


\pagerange{\pageref{firstpage}--\pageref{lastpage}} \pubyear{2012}

\maketitle

\label{firstpage}

\begin{abstract}
Microlensing observations indicate that quasar accretion discs have half-light
radii larger than expected from standard theoretical predictions based on
quasar fluxes or black hole masses. Blackburne and colleagues have also found a
very weak wavelength dependence of these half-light radii. We consider
disc temperature profile models that might match these observations.
Nixon and colleagues have suggested that misaligned accretion discs
around spinning black holes will be disrupted at radii small enough
for the Lense-Thirring torque to overcome the disc's viscous torque.
Gas in precessing annuli torn off a disc will spread radially and intersect
with the remaining disc, heating the disc at potentially large radii.
However, if the intersection occurs at an angle of more than a degree or so,
highly supersonic collisions will shock-heat the gas to a Compton temperature
of $T\sim 10^7$~K, and the spectral energy distributions (SEDs) of discs with
such shock-heated regions are poor fits to observations of quasar SEDs.
Torn discs where heating occurs in intermittent weak shocks that occur whenever
the intersection angle reaches a tenth of a degree pose less of a conflict with
observations, but do not have significantly larger half-light radii than
standard discs.
Our toy model for torn accretion discs is therefore unable to
simultaneously match observed SEDs and large half-light radii in quasars.
We also study two phenomenological disc temperature profile models.
We find that 
discs with a temperature spike at relatively large radii and lowered
temperatures at radii inside the spike yield improved and acceptable fits
to microlensing sizes in most cases.
Such temperature profiles could in principle occur
in sub-Keplerian discs partially supported by magnetic pressure.  However,
such discs overpredict the fluxes from quasars studied with microlensing
except in the limit of negligible continuum emission
from radii inside the temperature spike.
\end{abstract}


\section{Introduction} \label{intro}

Standard accretion disc theory
\citep[][hereafter SS73 and NT73, respectively]{ss73,1973blho.conf..343N}
predicts a disc's effective temperature as a function of radius,
from which its half-light radius as a function of wavelength can be calculated.
The predicted disc temperature profile is 
\begin{equation} \label{eqT}
T(R)\propto f_{\rm Edd}^{1/4} M_{BH}^{1/2} (R/R_g)^{-3/4}
\end{equation}
at $R\gg R_g$, where $R_g\equiv GM_{BH}/c^2$
with $M_{BH}$ the black hole mass and \fedd\ the 
quasar's bolometric luminosity relative to its Eddington luminosity.
The corresponding prediction for the variation of the half-light radius with
wavelength is 
\begin{equation} \label{eqR}
R_{1/2}\propto f_{\rm Edd}^{1/3} M_{BH}^{2/3} \lambda^{4/3} 
\end{equation}
for a broad range of wavelengths.

Quasar accretion disc half-light radii can be inferred using observations
of gravitationally lensed quasars \citep[see, e.g.,][]{2006glsw.conf..453W}.
Microlensing by stars in the lensing galaxy causes the flux ratios of the
lensed images to deviate from the predictions of models with smooth mass
distributions.
The amplitude of these deviations depends on the size of
the emission region at the wavelength being observed relative to the
typical Einstein radius of the lensing stars.

Observationally, it has been found that quasar accretion discs have
{\em microlensing sizes} (inferred from microlensing observations)
which are larger than the {\em theory sizes} \citep{2010ApJ...712.1129M}
predicted from theory given each quasar's estimated black hole mass
and assuming accretion at the Eddington rate, 
10\% accretion efficiency 
and an average disc inclination to the line of sight of 60$^\circ$.
In the three studies to date with sample sizes of ten or more objects,
the microlensing sizes range from 
a factor of $\sim 1.8\pm 1.6$ higher on average
at mean rest-frame wavelength 2660\,\AA\ \citep[hereafter M10]{2010ApJ...712.1129M}
to a factor of $\sim 5_{-4}^{+3}$ \citep{2012ApJ...751..106J} 
to $\sim 10^{+7}_{-5}$ 
\citep[][hereafter B11]{2011ApJ...729...34B}\footnote{To calculate this
factor for B11,
we use their median half-light radii computed with a logarithmic prior and
scale to the Eddington ratio and accretion efficiency used by other references.
}
higher on average at mean rest-frame wavelength 1736\,\AA.

The above results show that
microlensing sizes are larger than theory sizes at only marginal significance
(assuming accretion at the Eddington limit, which may be an overestimate;
see \citealt{2006ApJ...648..128K}).
However, microlensing sizes are significantly larger than quasar 
{\em flux sizes} (M10). 
Flux sizes are found by
determining how large a standard disc would need to be
to generate the specific luminosity corresponding to the
observed magnification-corrected flux of the quasar at a given wavelength.
M10 find that microlensing sizes are $\sim 4\pm 2$ times larger
than flux sizes.
Similar results have been obtained in studies of individual objects
\citep[e.g.,][]{2008ApJ...689..755M,2010ApJ...709..278D,2010ApJ...712..668P,2011ApJ...730...16M,2012ApJ...756...52M}.

As for the wavelength dependence of quasar half-light radii,
the microlensing sizes in the study of B11 are very weakly
dependent on wavelength in ten of eleven cases studied,
implying in the simplest case a temperature profile steeply
decreasing with radius at temperatures that generate ultraviolet emission.
On average,
B11 finds $R_{1/2} \propto \lambda^\nu$ with $\nu={0.17\pm 0.15 \pm 0.13}$
(quoting random and systematic errors), where $\nu=4/3$ is predicted by theory.

The results of B11 contrast with studies of individual objects which have
generally found results consistent with the theoretical prediction:
$\nu=1.64_{-0.42}^{+0.63}$ for HE 1104$-$1805 \citep{2008ApJ...673...34P};
$\nu=1.48_{-0.43}^{+0.60}$ for MG 0414+0534 \citep{2008MNRAS.391.1955B},
as compared to $\nu=1.50\pm 0.84$ for that object in B11;
$\nu=1.3 \pm 0.3$ for HE 0435$-$1223 \citep{2011ApJ...728..145M},
as compared to $\nu=0.67\pm 0.55$ for that object in B11;
$\nu=1.2 \pm 0.3$ for Q 2237+0305 \citep{2008A&A...490..933E};
$\nu=1.0 \pm 0.5$ for SBS 0909+532 \citep{2011ApJ...730...16M}; and
$\nu<4/3$ at 94\% confidence for SDSS J0924+0219 \citep{2009MNRAS.398..233F},
as compared to $\nu=0.17\pm 0.49$ for that object in B11.

Further studies of microlensed quasars are clearly needed
\citep{2011ApJ...738...96M}.
Meanwhile, investigation of possible explanations for unexpectedly large sizes
and nonstandard temperature profiles in quasar accretion discs is warranted.

To summarize:
there is marginal evidence that quasar accretion discs
are larger than their theory sizes 
(though the evidence strengthens 
if quasars are not typically accreting at the Eddington limit),
considerable evidence that they are larger than their flux sizes,
and possible evidence that their half-light radii have a different wavelength
dependence than that predicted by theory.  As discussed in M10,
a complication to the latter issue is that reconciling the three size
measurements may require {\em flatter} temperature profiles ($\beta < 3/4$) 
and thus a steeper size-wavelength relation ($\nu = \beta^{-1} > 4/3$).
The two constraints are not necessarily contradictory; a nonmonotonic 
temperature profile can be locally steep yet effectively flat overall.
If the correct temperature profile and physical parameters 
are used, theory sizes and flux sizes should agree with each other
and with the microlensing sizes and the observed wavelength dependence
of the sizes should match that predicted by the temperature profile.

Other options discussed in M10 to help explain the three discrepant size 
measurements include low accretion efficiency for unobscured quasars
or increasing the apparent disc size through 
contamination by line emission arising on larger physical scales
or through scattering \citep[][see also]{2010ApJ...709..278D}.  
For example, a large fraction of the disc continuum could be intercepted 
by a strongly warped disc \citep[e.g.,][]{2005MNRAS.359..545N,td13}.  
In that case,
the half-light radius will increase with a wavelength dependence related to the
disc albedo, as radiation not scattered by the disc will be absorbed
and reradiated at the local equilibrium temperature (which will be lower than
the characteristic temperature of the continuum emission from smaller radii).

Disc sizes also increase with Eddington ratio, so super-Eddington accretion 
may also help explain discrepant size measurements.
Furthermore, \citet{2012MNRAS.427.1867A} have suggested that 
the steep temperature profile found by B11 arises from the formation of 
a scattering photosphere in gas outflowing from the inner regions of 
an accretion disc undergoing super-Eddington accretion
\citep[see also][]{2013ApJ...770...30B,srm13}.
The small size of the X-ray emitting regions in quasars 
\citep[e.g.,][]{2012ApJ...756...52M}
could be explained in their model if the outflowing gas forms a 
Compton-thick funnel-shaped wind and photosphere, so that quasars
are only visible within the cone of the funnel because the funnel 
obscures both the X-ray and UV continuum source regions
when viewed from other angles.

Another possible explanation for the temperature profiles found by B11
may arise from changes to an
accretion disc resulting from misalignment of its initial angular momentum
vector and the angular momentum vector of a spinning black hole.

Any material in the vicinity of a massive spinning object experiences a
precession induced by the frame dragging due to the object \citep{1918PhyZ...19..156L}.
The result of this precession, known as the Lense-Thirring effect,
on a misaligned disc around the object is that each individual ring
in the disc will precess at different rate, warping the disc.

\citet{1975ApJ...195L..65B}
showed that the discs around spinning black holes evolve in such a
way that in the final configuration the disc is warped, with
the orbital angular momentum of the inner region either
aligned or anti-aligned with the spin axis
of the black hole while the disc's outer region remains unperturbed.
Warped-disc dynamics has been the subject of extensive research
\citep[e.g.,][]{1983MNRAS.202.1181P,1985MNRAS.213..435K,1992MNRAS.258..811P,1995ApJ...438..841P,1999MNRAS.304..557O,2002MNRAS.337..706L}.

In several recent papers, Nixon and colleagues have suggested that the
inner regions of inclined discs could
evolve by tearing apart instead of smoothly warping.

\citet[hereafter NKPF]{2012ApJ...757L..24N} 
showed that for realistic parameters, the
inner region of randomly oriented accretion discs around a spinning
black hole will tear, due to the Lense-Thirring effect.
If the angle between the outer disc and the black hole spin is in
the range $\sim 45^{\circ} -135^{\circ}$, the misalignment between
the inner edge of the
outer disc and the annulus (of width $\sim H$) torn off it
will be $> 90^{\circ}$ after half a precession
period.  The rotational velocities of those gas parcels will
then be partially opposed, leading to cancellation of
angular momentum and rapid infall \citep*[see][]{2012MNRAS.422.2547N}.
NKPF suggested that, given the smallness of the misalignment
angle upper limit for avoiding disc tearing (of order a few degrees),
the effect should be common in accretion around
the supermassive black holes that power quasars.

When gas in a precessing annulus torn from a disc intersects with gas in an
annulus still connected to the disc, the intersecting gas will be collisionally
heated.  High-temperature gas in intersection regions at relatively large
distances from the central black hole can in principle help explain the large,
relatively wavelength-independent half-light radii
inferred for quasar accretion discs.
In addition, the wide range of behaviors possible in the torn disc model
as a function of the initial disc-BH spin misalignment angle
could in principle help explain the wide range of wavelength dependences
in quasar accretion disc half-light radii \citep{2011ApJ...729...34B}.

In this paper we examine whether the temperature profiles of torn and other
non-standard discs can simultaneously reproduce the observed spectral energy 
distributions and inferred large half-light radii of quasar accretion discs.
In \S~\ref{toy} we outline the initial torn-disc model used in our study.
In \S~\ref{tprof} we discuss the temperature profiles and SEDs for
selected torn discs.
In \S~\ref{models} we explore a small set of additional temperature profiles.
We discuss our results in \S~\ref{discuss}.
Where necessary, we assume a flat universe with
$\Omega_M=0.3$, $\Omega_\Lambda=0.7$, and \ho\,=\,70\,\kmsm.

\section{Torn disc toy model} \label{toy}

Our initial model begins with accreting material forming a Keplerian disc at
large radii from the black hole.  The disc's spin axis is oriented at a random
angle $0^\circ < \theta < 180^\circ$ relative to the black hole spin axis.

The disc spreads inwards and different annuli in it precess
at different rates due to the Lense-Thirring effect,
so the disc warps as it spreads inwards.

The disc eventually reaches a radius where viscous dissipation cannot
overcome the shear between adjacent annuli imposed by the Lense-Thirring effect.

The innermost annulus of the disc then tears away
from the rest of the disc and precesses independently.
This annulus has width $\sim H$, where $H$ is the local disc height.

There will be two contact points between the precessing annulus and the next
innermost annulus (which forms the inner edge of the outer disc).
At these points gas from each annulus on intersecting orbits is likely to
shock, converting some orbital motion to heat.  The post-shock gas will be
unable to remain in circular orbits at the intersection radius.
The original NKPF picture consists of gas intersecting at angles
up to $2\theta$.

\subsection{Break radius and circularization radius} \label{radii}

The disc tears at a break radius $R_b$ given by Eq. 7 and Eq. 8 of NKPF:
\begin{equation} \label{Rb0}
R_b(\theta) \lesssim
\left(\frac{4}{3}|\sin\theta|\frac{a}{\alpha}\frac{R}{2H}\right)^{2/3} R_g
\end{equation}
\begin{equation} \label{Rb}
R_b(\theta) \lesssim 350 R_g |\sin\theta|^{2/3}
\left(\frac{a}{0.5}\right)^{2/3}
\left(\frac{\alpha}{0.1}\right)^{-2/3}
\left(\frac{2H/R_b}{10^{-3}}\right)^{-2/3} \end{equation}
where $2H$ instead of $H$ appears in the above equations because
we use $H$ to refer to the disc scale height, whereas NKPF 
use it to refer to the disc thickness.

Gas which precesses and intersects at $R_b$ loses angular momentum,
meaning that it can recircularize its orbit only at some 
circularization radius $R_{circ}<R_b$. \citet{2012MNRAS.422.2547N}
give an approximate expression for $R_{circ}$
(their Eq. 3), but the exact expression for the circularization radius
of an annulus of width $H$ centered at $R_b$$-$$H/2$
and an annulus of width $H$ centered at $R_b$$+$$H/2$,
intersecting at an angle $2\theta$ is
\begin{equation} \label{Rcirc}
\frac{R_{circ}(\theta)}{R_b(\theta)}=\frac{1}{2} + \frac{1}{2}\sqrt{1-\left(\frac{H}{2R_b}\right)^2}\cos(2\theta).
\end{equation}
Note that the above is the minimum possible $R_{circ}$;
if gas intersects at angle $\eta<2\theta$,
it will circularize at an intermediate radius $R_{circ} < R < R_b$.

\subsection{Shock heating}

When an annulus torn off a disc oriented at angle $\theta$ to
the black hole spin precesses, its intersection angle with the
adjacent annulus that forms the inner edge of the outer disc
will cycle sinusoidally between 0 and 2$\theta$.
The intersection angle between these annuli as a function of time,
$\eta(t)$, is given by
\begin{eqnarray} \label{e_eta}
\eta(t)=2\theta\times |\sin(180^\circ t/t_p)| & & 0^\circ<\theta<90^\circ \nonumber\\
\eta(t)=(360^\circ-2\theta)\times |\sin(180^\circ t/t_p)| & & 90^\circ<\theta<180^\circ
\end{eqnarray}
where the precession period is $t_p=360^\circ/\Omega_p$, with
$\Omega_p$ the Lense-Thirring precession frequency in degrees
per unit time.

At each intersection, each annulus approaches at an
angle $\eta/2$ to their mutual bisector.
We assume that the kinetic energy from motion transverse to the bisector
is converted into thermal energy.
This kinetic energy is
$K = 2 \times \frac{1}{2} (N \mu m_H) [v_K \sin (\eta/2)]^2$,
where $N$ is the number of particles (of mean mass $\mu m_H$)
in each annulus
and $v_K$ is the Keplerian velocity at the intersection radius.
If all this transverse kinetic energy is converted to heat,
the resulting thermal energy is distributed among the same $2N$
particles: $E = \frac{3}{2}(2N)k_b\Delta T$, where $\Delta T$ is
the temperature increase of the gas.
Thus, for gas intersecting at angle $\eta$ at radius $R_b(\theta)$,
\begin{equation} \label{DeltaT}
\Delta T = \frac{\mu m_H}{3k_b} \left[v_K(R_b(\theta)) \right]^2 \sin^2\frac{\eta}{2}.
\end{equation}
This $\Delta T$ forms an additional term for the surface temperature
$T_{\rm surf}$
in a region of width $H(R_b)$ wherever a precessing annulus is present.

If gas is shock-heated to a temperature higher than the Compton temperature
of the ambient radiation field, the gas will cool by transferring its
thermal energy to the radiation field through inverse Compton scattering.
The Compton temperature of a quasar radiation field is $\sim 10^7$~K
\citep[Eq. 25 and 26 of][]{2004MNRAS.347..144S}.
We therefore cap the value of $T_{\rm surf}$ at $10^7$~K.

\section{Torn disc toy model SEDs and half-light radii} \label{tprof}

The temperature profile and SED of a torn disc ultimately depend on
the disc/BH spin misalignment angle $\theta$,
the dimensionless BH spin parameter $a$ ($0\leq a \leq 1$),
the dimensionless viscosity parameter $\alpha$,
the disc height to radial distance ratio $H/R$,
and the black hole mass $M_{BH}$ and normalized mass accretion rate
$\dot{M}c^2/L_{\rm Edd} \propto \dot{M}/M_{BH}$.
B11 write this latter quantity as $f_{\rm Edd}/\eta$, where
the Eddington ratio is
$f_{\rm Edd}=L_{bol}/L_{\rm Edd} =\dot{M}/\dot{M}_{\rm Edd}$
and the accretion efficiency is $\eta=L_{bol}/\dot{M}c^2$.
B11 assumed $\eta=0.15$ and $f_{\rm Edd}=0.25$,
whereas in all cases we assume a Schwarzschild black hole ($\eta=0.057$).
To match the value of $f_{\rm Edd}/\eta$ used by B11,
we assume $f_{\rm Edd}=0.1$.
We also assume $H/R=1.7\times 10^{-3}$ (so that the disc height 
at any radius is $H=1.7\times 10^{-3}R$), $\alpha=0.1$,
and $M_{BH}=10^9~M_\odot$. 
In our illustrative calculations below, we adopt a SS73 temperature profile
(valid only for $a=0$) even for the case of $a\neq 0$, as 
our arguments are insensitive to the differences
between SS73 and NT73 SEDs.

\subsection{Single-$R_b$ case} \label{singleRb}

Equations \ref{Rb}, \ref{Rcirc} and \ref{DeltaT}
generate a family of temperature profiles as a function of $\theta$
which consist of:

$\bullet$~a normal SS73 disc ($T_{\rm surf}=T_{\rm SS73}$) at $R<R_{circ}(\theta)$,

$\bullet$~a gap at $R_{circ}<R<R_b(\theta)-H(R_b)$,

$\bullet$~a shock-heated region
($T_{\rm surf}=T_{\rm SS73}+\Delta T$) at radii $R_b-H(R_b)<R<R_b+H(R_b)$,

$\bullet$~and a normal SS73 disc again at $R>R_b+H(R_b)$.

To find the temperature profile for the scenario put forth by NKPF,
in which the precessing disc intersects with its neighboring
annulus over half to one precession periods,
for a given $\theta$ we
use Eq. \ref{DeltaT} to calculate $T_{\rm surf}(R_b)=T_{\rm SS73}(R_b)+\Delta T(R_b)$
for $\eta=\theta$ ($\theta\leq 90^\circ$)
or $\eta=180^\circ-\theta$ ($\theta\geq 90^\circ$), as the average of $\eta$
over one precession period is about half the maximum $\eta$.

We then calculate spectral energy distributions (SEDs)
and half-light radii using the $L_\lambda$ values
from the regions $R<R_{circ}$, $R_b-H<R<R_b+H$ and $R>R_b+H$.

\begin{figure*}
\includegraphics[angle=0, width=0.670\textwidth]{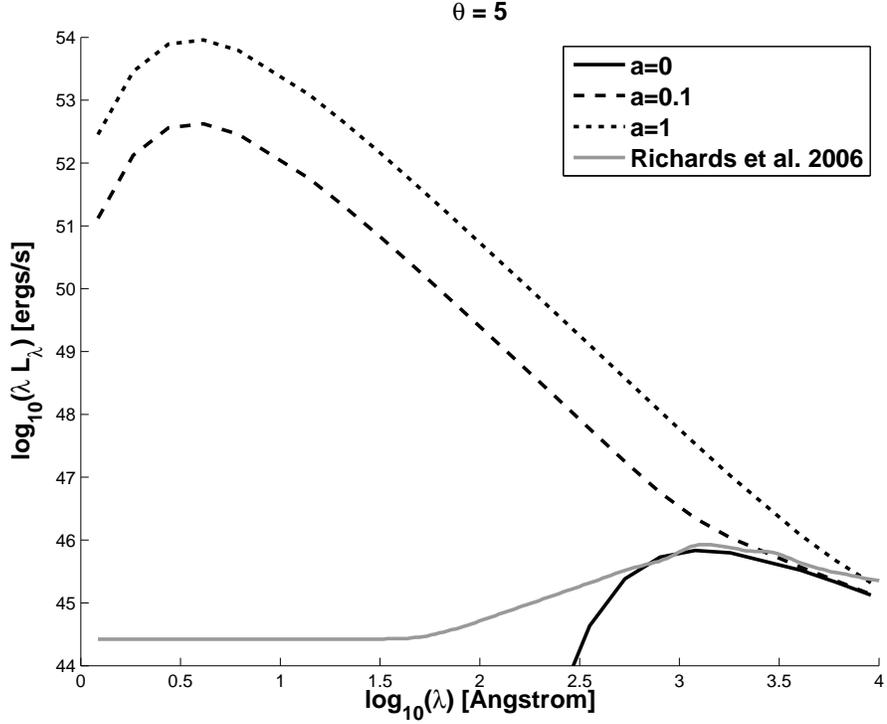}
\caption{SEDs of accretion discs with $M_{BH}=10^9~M_\odot$ and 
\fedd=0.1.  
The solid black line is the SED of an accretion disc with $a=0$.
The dashed and dotted lines show the SEDs of discs tilted by $\theta=5^\circ$ 
from the spin axis of black holes with $a = 0.1$ and 1, respectively, that tear
and shock-heat to $T = 10^7$~K at $R_b$ over an annulus of width $2H(R_b)$.
The grey line is the mean quasar SED from \citet{gtr06}.
}
\label{fig:1} \end{figure*}

Figure \ref{fig:1} shows an SS73 disc SED
and the SEDs resulting from these shock-heated disc temperature profiles.
The SEDs are much bluer and more X-ray luminous than observed quasar SEDs.
(Note that the SS73 disc model is known to be a poor fit to observations
at $\lambda \lesssim 1000$~\AA, and that at X-ray wavelengths an additional
power-law component is needed in addition to thermal disc emission.
Those wavelengths are plotted here to examine if an SED overpredicts the
short-wavelength emission, not in expectation of a good fit at short
wavelengths from any SED, even a baseline SS73 one.)

\begin{figure*}
\includegraphics[angle=0, width=0.670\textwidth]{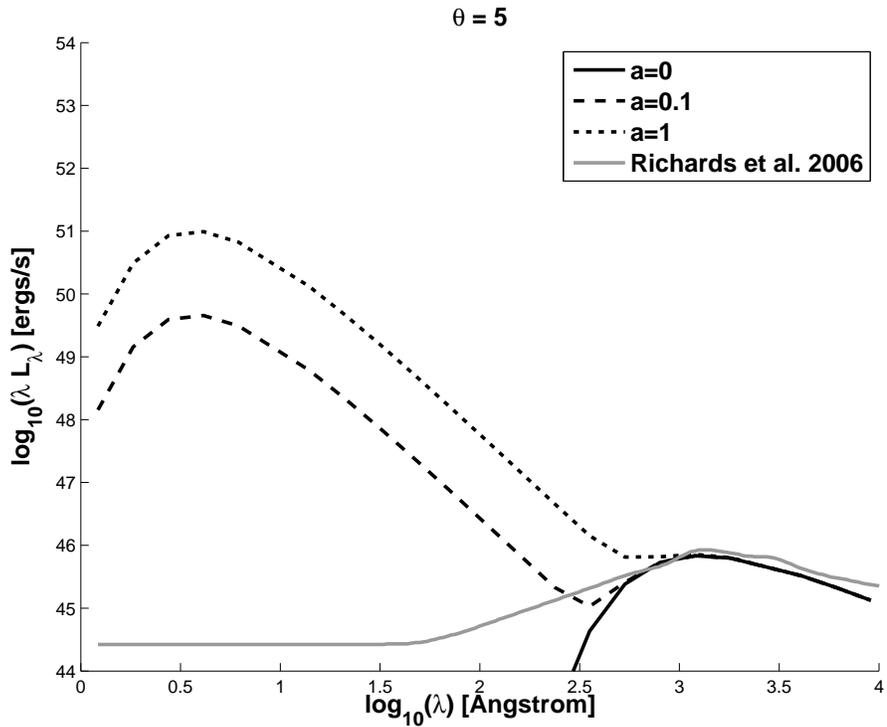}
\caption{Same as Figure \ref{fig:1}, except that shock heating occurs 
only in the intersecting regions of the torn annulus and 
the adjacent annulus that forms the inner region of the outer disc.}
\label{fig:2} \end{figure*}

We have assumed that an entire annulus at $R_b$ is at the post-shock
temperature, 
but it may be more accurate to assume that only the two intersection
regions of surface area $(2H)^2$ on each face of the disc are at that
temperature.  Figure \ref{fig:2} shows the SEDs in that case; they
are still much bluer and more X-ray luminous than observed quasar SEDs.

We discuss the half-light radii results for this and the next model
in \S~\ref{toyresults}.

\begin{figure*}
\includegraphics[angle=0, width=0.670\textwidth]{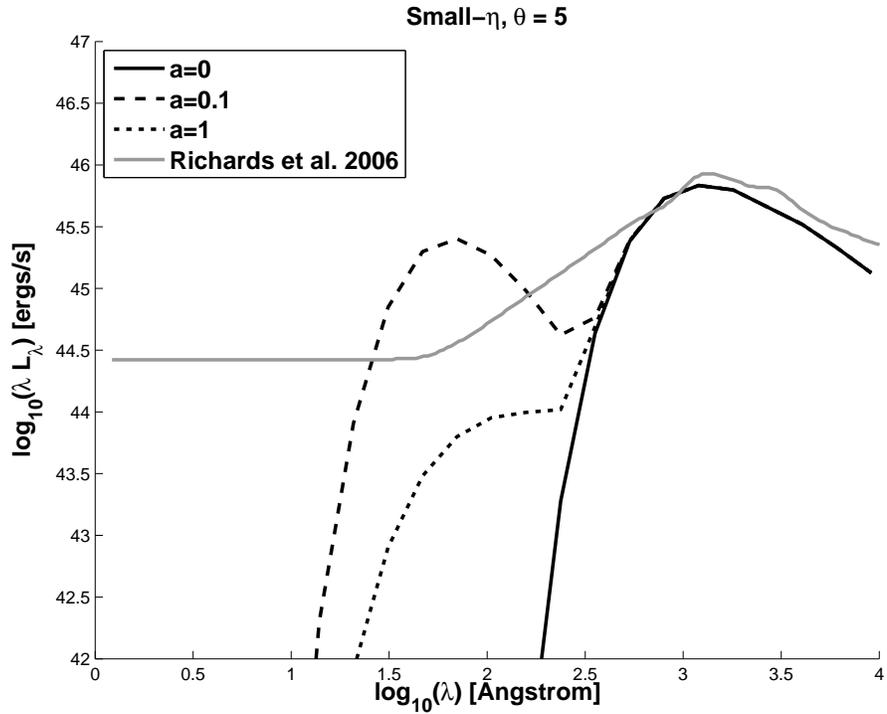}
\caption{The solid black line is the SED of an accretion disc with $a=0$.
The dashed and dotted lines show the SEDs of discs tilted by $\theta=5^\circ$
from the spin axis of black holes with $a = 0.1$ and 1, respectively, 
that tear and shock-heat in multiple, short-lived shocks, each with
intersection angle $\eta=0.1^\circ$.
See \S~\ref{smalleta} for a discussion of why the $a=1$ curve
and not the $a=0.1$ curve is most similar to the $a=0$ curve.
The grey line is the mean quasar SED from \citet{gtr06}.}
\label{fig:3} \end{figure*}

\begin{figure*}
\includegraphics[angle=0, width=0.670\textwidth]{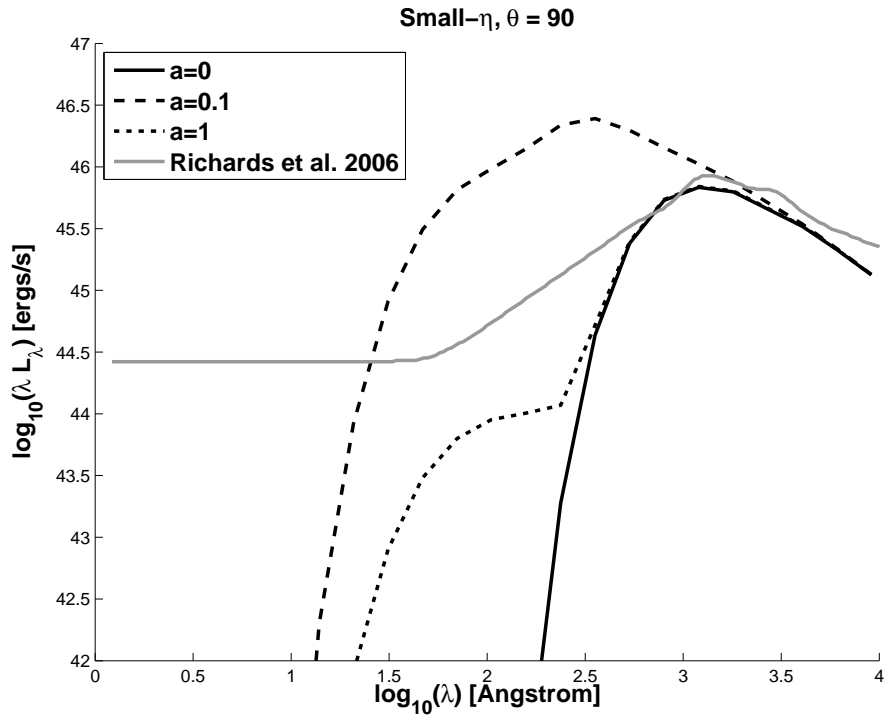}
\caption{Same as Figure \ref{fig:3}, except for $\theta=90^\circ$.}
\label{fig:4} \end{figure*}

\setcounter{table}{0}
\begin{table*}
 \centering
 \begin{minipage}{85mm}
 \caption{Model accretion disc half-light radii at $\lambda$=1736~\AA}
  \begin{tabular}{@{}llccc@{}}
\hline
Model        & Shocked Region     & $a$ & $\theta(^\circ)$ & $\log(r_{1/2}{\rm ~in~cm})$ \\
\hline
Inferred (B11)         & ...                & ...                & ... & 15.97$-$16.61 \\ 
SS73                   & None               & 0                  & ...      & 15.55    \\
Single-$R_b$ torn disc & Full Annulus       & 0.1                & 5        & 15.19    \\
Single-$R_b$ torn disc & Full Annulus       & 0.1                & 90       & 15.90    \\
Single-$R_b$ torn disc & Full Annulus       & 1                  & 5        & 15.86    \\
Single-$R_b$ torn disc & Full Annulus       & 1                  & 90       & 16.56    \\
Single-$R_b$ torn disc & Intersections Only & 0.1                & 5        & 15.55    \\
Single-$R_b$ torn disc & Intersections Only & 0.1                & 90       & 15.98    \\
Single-$R_b$ torn disc & Intersections Only & 1                  & 5        & 15.56    \\
Single-$R_b$ torn disc & Intersections Only & 1                  & 90       & 16.56    \\
Small-$\eta$ torn disc & Intersections Only & 0.1                & 90       & 15.64    \\
Small-$\eta$ torn disc & Intersections Only & 0.3                & 90       & 15.59    \\
Small-$\eta$ torn disc & Intersections Only & 1                  & 90       & 15.56    \\
\hline
\end{tabular}\\
The B11 range of $\log(r_{1/2})$ comes from the four objects with masses closest to 
$M_{BH}=10^9$ $M_\odot$.
SS73, single-$R_b$ and small-$\eta$ models assume $M_{BH}=10^9$ $M_\odot$ and \fedd=0.1.
\label{tab:1}
\end{minipage}
\end{table*}

\subsection{Small-$\eta$ case} \label{smalleta}

Large intersection angles between the precessing annulus and the outer disc
lead to high-temperature shock-heated regions and SEDs which are inconsistent
with observations.  A more plausible alternative picture is one in which the
intersection angle never becomes large.

As the annulus at $R_{b}$ starts precessing, the intersection angle between it
and the next annulus out will grow with time.  The relative velocity between
those annuli will also grow with time.  Shock heating between the two annuli
will happen once the magnitude of 
their relative velocity is sufficiently large, rather than
at the maximum relative velocity possible for that value of $\theta$.
Therefore, the relative velocity at the time of intersection
may not depend on $\theta$,
and the shock-heated temperature may be considerably lower than $10^7$~K.

The intersection angle between these two annuli as a function of time,
$\eta(t)$, is given by Equation \ref{e_eta}.
The magnitude of the relative velocity between two annuli at $R_b$
divided by the speed of sound at $R_b$, $c_s=(H/R_b)v_K$, 
is given, for $0^\circ<\theta<90^\circ$, by
\begin{eqnarray} \label{e_vrel}
\frac{v_{rel}(R_b)}{c_s(R_b)} = \frac{2 v_K(R_b)}{[H/R_b]v_K(R_b)} \sin[\frac{1}{2}\eta(t)] = \frac{2R_b}{H} \sin[\frac{1}{2}\eta(t)]
\end{eqnarray}
Setting $v_{rel}/c_s=1$ and solving, we find that the value of $\eta$ at which
intersecting annuli will shock is $\eta\simeq 0.1^\circ$.

We now imagine a different scenario for a disc oriented at $\theta$.
An annulus breaks off at $R_b(\theta)$, precesses by $\eta=0.1^\circ$, shocks,
and reorients itself and its adjacent annulus to $\theta-0.05^\circ$.
Part of this reoriented annulus will be subject to breaking at
$R_b(\theta-0.05^\circ)$, as the separation between the two break radii will be
less than $H$ for most $\theta$.

We suppose that material flows to smaller radii through a number
$\theta/0.05^\circ$ of these weak shocks.
Each shock heats the disc in two intersection regions of surface area $(2H)^2$
on each face of the disc.
Each pair of regions will form at some radius $R_b(\theta_x)$ 
at the inner edge of part of the disc oriented at some angle $\theta_x$, 
and the disc material will
flow through one or the other of the regions in half an orbital time.
Thus, each shock exists only for a time of $\frac{1}{2}t_{orb}/t_{rad}$,
where $t_{rad}$ is the time it takes material to flow from
$R_b(\theta_x)$ to $R_b(\theta_x-0.05^\circ)$ in an SS73 disc.
We compute the average SED expected for such a disc by setting the shock-heated
regions' radial widths to $w(R)$, where
$2\pi R w \equiv 2(2H)^2 \times \frac{1}{2}t_{orb}/t_{rad}$.
The shock-heated temperature in each region is
\begin{equation} \label{deltaTx}
\delta T(\theta_x) = \frac{\mu m_H}{3k_b} \left[v_K(R_b(\theta_x)) \right]^2 \sin^2 (0.05^\circ).
\end{equation}

The average SEDs of discs in this ``small-$\eta$'' scenario are shown in
Figure \ref{fig:3} for the case $\theta=5^\circ$ and in
Figure \ref{fig:4} for the case $\theta=90^\circ$.
In both cases, the average SEDs for $a=1$ 
are consistent with observational constraints,
while those for $a=0.1$ are not consistent at sufficiently large $\theta$.
This somewhat counterintuitive behaviour arises mainly from the fact that 
$R_b$ increases with $a$ (Eq. \ref{Rb}).
In the small-$\eta$ scenario, shock locations are 
at smaller radii and are more closely spaced for $a=0.1$ than for $a=1$.
Because both the shock-heating temperature increment
$\delta T$ and $T_{\rm surf}=T_{\rm SS73}+\delta T$ increase with
decreasing radius and we assume the disc is optically thick ($L\propto T^4$),
the shock-generated luminosity is larger for smaller $a$.
(This increase does stop at very small $a$,
at which large values of $\theta$ are required for shocks to occur at radii
larger than the innermost stable circular orbit at which a disc can exist.)

Note that the SEDs in the small-$\eta$ scenario could in principle
be consistent with the range of observed quasar SEDs 
if most black holes are rapidly spinning, so that cases with $a\gg 0.1$
in Figures \ref{fig:3} and \ref{fig:4} were much more common than cases
with $a\lesssim 0.1$.
(The first test of that possibility would be to see if the SEDs are relatively
unchanged when calculated assuming an NT73 disc with the appropriate value of
$a$ instead of an SS73 disc.)
However, the SEDs predicted for low spins ($a \approx 0.1$; consistent with 
the accretion picture of \citealt{2006MNRAS.373L..90K,2007MNRAS.377L..25K})
do show a soft X-ray component which has recently been shown to be 
ubiquitous in type 1 AGN \citep{2012MNRAS.423.2633S}.

\subsection{Torn disc toy model results} \label{toyresults}

Table \ref{tab:1} gives the half-light radii 
at $\lambda=1736$~\AA\ inferred by B11 (microlensing sizes)
and calculated for the scenarios considered in this section (theory sizes),
all for quasars of mass $M_{BH} \simeq 10^9~M_\odot$.

Most torn disc cases have larger half-light radii than the SS73 case.
However, only the single-$R_b$, $a=1$, $\theta=90^\circ$ torn disc cases
(with a shocked region over the full annulus or at the intersections only)
have half-light radii large enough to match the
observationally inferred results of B11.
Within the uncertainties, single-$R_b$ torn discs with full-annulus
shocked regions around black holes with $a=1$ might match the observed 
range of $r_{1/2}$ for a reasonable distribution of angles $\theta$.
However, those scenarios do not produce SEDs consistent with observations.
And while the small-$\eta$ scenarios are consistent with 
observed SEDs for certain parameter choices, none of those scenarios 
result in half-light radii large enough to match the observations
when averaged over a reasonable range of $\theta$ values.

These conclusions hold if \fedd\ is increased from 0.1 to 1.0.
In that scenario 
the temperature of the disc will be higher at all radii (Eq. \ref{eqT}),
increasing $\log(r_{1/2})$ by 0.33 in the SS73 case (Eq. \ref{eqR}),
which reduces but does not eliminate the size discrepancy.
However, because $H/R$ increases with \fedd\ (SS73 Eq. 2.8), the values 
of $R_b$ for torn discs will shrink as \fedd\ increases (Eq. \ref{Rb}).
That will lead to smaller values of $\log(r_{1/2})$ for torn discs
in cases where the shocked region dominates the emitted flux and values of
$\log(r_{1/2})$ only 0.33 higher in most other cases.

\section{Phenomenological model tests}	\label{models}

In this section 
we consider whether better matches to observations can be obtained 
with two phenomenological models for disc temperature profiles
that do not involve torn discs.
These models are energetically plausible but are not rigorously derived;  
our intention is simply to explore how well simple combinations of 
an underlying disc and a temperature spike can match observations.

We consider the constraints placed on these two phenomenological models 
by the quasar accretion disc half-light radii inferred by B11 
(microlensing sizes; see \S~1). 
B11 characterize the half-light radius' dependence on wavelength 
in their sample as $r_{1/2}=\lambda^\nu$, with $\nu = 0.17 \pm 0.15$.
The quantity $\nu$ is related to the slope $\beta$ of a disc temperature
profile $T(R)\propto R^{-\beta}$ as $\beta=\nu^{-1}$;
thus, B11 constrain $\beta \simeq 6_{-3}^{+\infty}$.

The weak dependence of the half-light radii on wavelength seen by B11
extends over the wavelength range $0.1 - 1 ~\micron$, corresponding to
wavelengths of light emitted by blackbodies at temperatures 
of approximately $3,000~K < T < 30,000~K$.

Therefore, one simple explanation for the results of B11
is that quasar accretion discs 
contain a region where the temperature drops 
from $T\gtrsim 30,000$~K to $T\lesssim 3,000$~K
following a power-law of the approximate form 
$T(R)\propto R^{-\beta}$ with $\beta\simeq (6_{-3}^{+\infty})$.\footnote{Other 
explanations are possible.  
The locally inhomogeneous accretion disc model of \cite{da11} explains
large half-light radii, but not their lack of wavelength dependence.
A modified inhomogeneous disc model in which regions of the disc flare to some 
constant temperature $T\gtrsim 30,000$~K within a relatively large radius,
but not outside that radius, might explain both trends.
However, as we can see no obvious mechanism for generating a radially
constant peak flare temperature less than the Compton temperature,
we have chosen to explore other models here.}

We parametrize our models in terms of $M_{BH}$ (fixed for each quasar),
the Eddington ratio \fedd (for which we test a wide range of values),
and other parameters appropriate to each of the two models we now present.

\begin{table*}
 \centering
 \begin{minipage}{\linewidth}
 \caption{Tabulated and derived parameters for quasars from Blackburne et al. (2011)}
  \begin{tabular}{@{}lllccccccccr@{}}
\hline
            &                 &	$L_{bol,opt}$              & $\sigma(\log$  & Measured                      &             & SS73                          & $\log(r_I)$, & $\log(r_I)$, &                   \\ 
            & $M_{BH}$        & ($10^{46}$                 & $L_{bol,opt}$) & $\log\langle r_{1/2} \rangle$ & $\log(r_Q)$ & $\log\langle r_{1/2} \rangle$ & $\beta=3/4$  & MR best-fit  & $\sigma_r/\sigma$ \\ 
Quasar      & $(10^9M_\odot)$ &            erg~s$^{-1}$) & ($\log$ erg~s$^{-1}$) & $(\rm cm)$               & (cm)        & $(\rm cm)$                    &  (cm)        &  (cm)        &                   \\ \hline 
HE 0230$-$2130     & 0.092  & 0.29 & 0.24 & 16.57 & 16.76   & 15.13 & 14.73 & 16.73 & 0.148 \\ 
MG J0414$+$0534    & 1.82   & 3.6  & 0.17 & 15.94 & 16.81   & 16.05 & 14.28 & 14.38 & 0.278 \\ 
HE 0435$-$1223     & 0.50   & 0.38 & 0.26 & 16.07 & 16.79   & 15.72 & 14.95 & 15.11 & 0.390 \\ 
RX J0911$+$0551    & 0.80   & 1.3  & 0.18 & 16.19 & 16.80   & 15.69 & 14.86 & 15.12 & 0.324 \\ 
SDSS J0924$+$0219  & 0.11   & 0.06 & 0.56 & 15.73 & 16.77   & 15.30 & 14.64 & 14.88 & 0.530 \\ 
HE 1113$-$0641     & 0.087  & 0.27 & 0.26 & 15.82 & 16.76   & 15.34 & 14.76 & 14.86 & 0.997 \\ 
PG 1115$+$080      & 1.23   & 1.1  & 0.37 & 16.69 & 16.81   & 15.99 & 15.21 & 15.59 & 0.626 \\ 
RX J1131$-$1231    & 0.06   & 0.08 & 0.19 & 15.51 & 16.76   & 15.36 & 14.88 & 14.94 & 0.342 \\ 
SDSS J1138$+$0314  & 0.04   & 0.38 & 0.26 & 15.98 & 16.75   & 14.83 & 14.64 & 14.90 & 0.944 \\ 
SDSS J1330$+$1810  & 1.5    & 4.7  & ...  & 16.13 & 16.81   & 16.12 & 14.24 & ...   & ...   \\ 
WFI J2026$-$4536   & 0.79   & 2.5  & 0.26 & 16.42 & 16.80   & 15.77 & 15.03 & 15.29 & 0.474 \\ 
WFI J2033$-$4723   & 0.18   & 0.57 & 0.12 & 16.69 & 16.77   & 15.42 & 14.90 & 15.16 & 0.508 \\ 
\hline
\end{tabular}\\
Values of $M_{BH}$ and $L_{bol,opt}$ are taken from Table 8 of B11.
Uncertainties on $L_{bol,opt}$ are discussed in \S~\ref{random}.
The measured $\log \langle r_{1/2} \rangle$ column gives the 
average of each object's log-prior, median $r_{1/2}$ values from B11.
The $r_Q$ values are discussed in \S~\ref{fd}.
The SS73 $\langle r_{1/2}\rangle$ calculations used $f_{\rm Edd}=0.1$
to match B11 given our assumed value of $\eta$ (see \S~\ref{tprof}). 
The $r_I$ flux sizes are discussed in \S~\ref{results}, for both the
$\beta=3/4$ case and the best-fit magnetically restrained (MR) model.
The $\sigma_r/\sigma$ values are discussed in \S~\ref{random}.
\label{tab:2}
\end{minipage}
\end{table*}

\subsection{Forming disc model} \label{fd}

One possibility for generating higher temperatures at larger radii
might be a fragmented disc transitioning to a continuous disc.
Goodman (2003) has argued that quasar accretion discs are likely to be 
unstable against fragmentation due to self-gravity at large radii.
If the gas in such a disc is distributed in clumps over some range of radii,
random collisions between those clumps will occur at higher velocities 
at smaller radii within that range.
The clumps will survive only down to a radius 
at which collisions disrupt them enough to establish a continuous disc, and
a temperature spike due to collisional heating is plausible near that radius.
This radius must be smaller than the maximum radius at which the disc is
stable against fragmentation, $r_Q$.
\begin{equation}
r_{Q} \simeq 415~ \alpha_{0.01}^{14/27} (f_{\rm Edd}/\eta)^{-8/27} 
\hat{\kappa}^{2/9} M_8^{-26/27}
R_{Sch}
\end{equation}
where $1 \lesssim \hat{\kappa}\lesssim 10$ 
is the opacity relative to the electron scattering opacity
and $M_8$ is the black hole mass in units of $10^8~M_\odot$.
We assume $\alpha_{0.01}=10$ and $\hat{\kappa}=10$, yielding
\begin{equation} \label{rQ}
r_{Q} \simeq 2292~ (f_{\rm Edd}/\eta)^{-8/27} M_8^{-26/27} R_{Sch}.
\end{equation}

In Table \ref{tab:2}, for each quasar in B11
we compare its value of $r_Q$ 
(assuming $f_{\rm Edd}/\eta=5/3$, to match B11)
to its measured $\langle r_{1/2} \rangle$,
where the average is over all wavelengths at which $r_{1/2}$ is measured by B11.
The values of $r_Q$ are all larger than the inferred $\langle r_{1/2}\rangle$,
indicating that a model with a temperature spike at $r\simeq r_Q$
might be able to match the observations.

We refer to this as our `forming disc' model, in which a smooth disc is 
continuously forming out of a fragmented disc at $R_b$.  
(We do not mean to imply that the disc is necessarily young.)  
If we denote the temperature profile of an SS73 disc as $T_{SS73}(R)$
then for the same parameters
the temperature profile of the forming disc model $T_{fd}(R)$ is:
\begin{eqnarray}
    T_{fd}(R<R_b) &=& T_{SS73}(R) \nonumber \\
T_{fd}(R\geq R_b) &=& T_{hi}(R/R_b)^{-\beta}.
\end{eqnarray}
For each quasar, this model
has one fixed parameter (\mbh) and four free parameters:
\fedd, the spike temperature $T_{hi}$, the spike radius $R_b$ 
(at which the temperature jumps to $T_{hi}$ from some temperature $T_{lo}$),
and the slope $\beta$ with which the temperature declines at $R>R_b$.

The values these parameters can take are constrained by energy conservation.
The energy per second emitted from the disc in its forming region ($R>R_b$)
must be no more than the energy available to the mass per second which has
reached that radius from infinity.
Consider the limiting case of the forming disc model consists of gas falling
radially inward in the disc plane until it reaches $R_b$, at which point it
circularizes onto the inner disc with kinetic energy per unit mass
$\frac{1}{2}v^2=\frac{1}{2}GM_{BH}/R_b$.
The potential energy per unit mass gained by reaching $R_b$ is $GM_{BH}/R_b$.
Thus, the forming region of the disc ($R>R_b$) can emit energy at a rate
$L\leq\frac{1}{2}G\dot{M}M_{BH}/R_b$,
with the characteristic width and peak temperature
of the extraction region dependent on $\beta$.

In appendix \ref{fdcon}, we use the above constraint to derive an expression
relating the free parameters of this model for $\beta>\frac{1}{2}$:
\begin{equation} \label{fdeq}
\sigma_{sb} T_{hi}^4 \leq \left[\beta-\frac{1}{2}\right] \frac{GM_{BH}\dot{M}}{2\pi R_b^3}
\end{equation}
which is in addition to the constraint $R_b \leq r_Q$, with $r_Q$ given by 
Equation \ref{rQ}).
We only consider combinations of parameters that satisfy those constraints.

For this model we test values of 
$\beta$ from 1 to 18,
$f_{\rm Edd}$ from 0.01 to 1,
$ T_{hi} $ from  $10000$~K to $70000$~K,
and $R_b = R(T_{lo})$ 
with $T_{lo}$ from 4000~K $<$ $T_{lo}$ $<$ 28000~K.
We detail our results on this and the next model in \S~\ref{results}.

\subsection{Magnetically restrained disc models} \label{mr}

A disc will have greater viscous dissipation (and thus higher temperatures)
than a Keplerian disc if gas flowing inward through a disc has 
a rate of change of azimuthal speed with radius 
whose absolute magnitude is greater than that experienced in a Keplerian disc.
If the magnitude of the
rate of change in azimuthal speed is less than in the Keplerian case,
the annulus will have a lower temperature than in the Keplerian case.

Both situations can occur (at different radii) in discs where magnetic
pressure is important. 
\cite{1997MNRAS.288...63O} has shown that a disc threaded by a strong
poloidal magnetic field can have an angular velocity with order unity
deviations from the Keplerian value.
In the magnetically arrested disc model of \cite{2003PASJ...55L..69N}, 
a strong poloidal magnetic field disrupts disc accretion at some radius,
inside of which gas accretes at much less than the free-fall velocity.
We consider a range of `magnetically restrained' models where 
gas accreting through
a Keplerian disc at large radii reduces its orbital velocity to a fraction
$f<1$ of the circular velocity at $R_b$ by the time it reaches radius $R_b$
(increasing the local disc temperature in the process)
and continues to orbit at a fraction $f$ of the Keplerian velocity at $R<R_b$.
(Sub-Keplerian rotation is required in the inner region to ensure
that gas moves inward at $R_b$ instead of being flung outward.)
Assuming that the radial velocities in the disc at $R<R_b$
are still much less than the rotational velocities,
the sub-Keplerian rotation reduces the dissipational heating at $R<R_b$,
leading to temperatures at $R<R_b$ which are $\sqrt{f}$ 
times the temperatures in an SS73 disc with the same parameters.
The combined profile has a cooler inner disc and warmer outer disc than in the
forming disc model.

Again denoting the temperature profile of an SS73 disc as $T_{SS73}(R)$,
for the same parameters
the temperature profile of a magnetically restrained disc model $T_{mr}(R)$ is:
\begin{eqnarray}
T_{mr}(R<R_b) &=& \sqrt{f}T_{SS73}(R) \nonumber \\
T_{mr}(R_b\leq R\leq R_c) &=& T_{hi}(R/R_b)^{-\beta} \nonumber \\
T_{mr}(R\geq R_c) &=& T_{SS73}(R)
\end{eqnarray}
where $R_c$ is defined via $T_{hi}(R_c/R_b)^{-\beta}\equiv T_{SS73}(R_c)$.
For each quasar, this model has one fixed parameter, $M_{BH}$, and five free
parameters: 
\fedd, $T_{hi}$, $R_b$ (or, equivalently, $T_{lo}$), $\beta$ and $f$.

The values of these parameters are constrained by the fact that
the disc has a fraction $(1-f^2)$ of the kinetic energy of the
accreting gas at $R_b$ available to emit as heat at $R>R_b$, in addition
to the normal disc emission ($\propto T_{\rm SS73}^4$) at $R>R_b$.
We derive the relationship between the parameters of this model in
appendix \ref{mrcon} (equation \ref{eqmrcon}), and require 
all combinations of parameters we test to satisfy that relationship.

For these magnetically restrained models we test the same parameter space of 
\fedd, $T_{hi}$, $R_b$ and $\beta$ used for the forming disc model,
plus values of $f$ given by $f=0$ and $f=0.01n^2$ for $n=2$ to $n=9$.
We detail our results on this and the previous model in \S~\ref{results}.

\subsection{Random and systematic errors} \label{random}

For each quasar in B11, for a specific model and set of model parameters
including that quasar's $M_{BH}$
we calculate half-light radii at the observed wavelengths
and a bolometric luminosity at $\lambda < 1~\micron$.
We then calculate the $\chi^2$ value of the model using
the model and observed $\log L_{bol,opt}$ values from \cite{2007ApJ...661...19P}
and B11 and the model and inferred half-light radii.
Most quasars have their half-light radius measured at eight wavelengths;
thus, the half-light radius has eight times greater weight in the $\chi^2$
calculation than $L_{bol,opt}$ does.

Inspection of Fig. 6 of B11 shows that the error bars
for a given object are generally larger than the scatter between points.
Using those error bars would
lead to misleadingly small $\chi^2$ and $\chi^2_\nu$ values.
The $\sigma$ values used in Fig. 6 of B11 incorporate both
random and systematic error, 
added together in quadrature as $\sigma=\sqrt{\sigma_r^2+\sigma_s^2}$.
Adding random and systematic errors together can lead to erroneous
results, but is likely an adequate approximation in in this case.
The systematic errors considered by B11 were
apparent flux ratio deviations arising from
emission-line contamination of the observed magnitudes,
confusion with adjacent objects on the images,
and variability from time delays.
Only the latter source of uncertainty is likely to systematically bias
the half-light radii at all wavelengths in the same direction;
emission-line contamination will vary between filters,
and confusion with adjacent objects will vary with the color of the objects.
While we first consider only $\sigma_r$ in our $\chi^2$ calculations,
we also show the results of a conservative approach
in which we consider the full $\sigma$.

To calculate the $\sigma_r$ values, 
we assume that each quasar's total error in $r_{1/2}$ in each filter
(Table 7 of B11) can be broken down into random and systematic errors,
in proportion to the contribution of random and systematic errors
to the uncertainties for all data points for that quasar (Table 5 of B11).
From the ratio $\sigma_r/\sigma_s$ we determine the ratio $\sigma_r/\sigma$
for each quasar (Table \ref{tab:2}).
The $r_{1/2}$ error values used in our $\chi^2$ calculations
are the quasar's uncertainties $\sigma$ from Table 7 of B11 
multiplied by the quasar's ratio $\sigma_r/\sigma$ from Table \ref{tab:2}.

The $\log L_{bol,opt}$ error value used in our $\chi^2$ calculations,
$\sigma(\log L_{bol,opt})$, 
is also tabulated for each quasar in Table \ref{tab:2}.
For quasars in \cite{2007ApJ...661...19P}, the $\sigma(\log L_{bol,opt})$ 
values here equal the $\sigma(\log M_{BH})$ there; otherwise, we used the 
average $\sigma(\log M_{BH})=0.26$ from \cite{2007ApJ...661...19P}
as a representative error on $\log L_{bol,opt}$.
We do not attempt to separate random and systematic errors on $L_{bol,opt}$.

\begin{table*}
 \centering
 \begin{minipage}{\linewidth}
 \caption{$\chi^2_{min}$ results}
  \begin{tabular}{@{}lcccc@{}}
\hline
Quasar             & Parameter              & SS73                 & Forming Disc          & Magnetically Restrained ($f$ value, $\kappa_\lambda$ value) \\ \hline
HE    0230$-$2130  & $\chi^2_{min}$         & 6176.1               & 6196.9                & 235.5 ($f=0$, $\kappa_\lambda=0.0001$)                       \\
                   & $f_{\rm Edd}$          & 1                    & 1                     & 1                                   \\
                   & $L_{bol,opt}$ ($10^{46}$ erg s$^{-1}$)        & 1.051                & 1.057                 & 0.003                               \\      
                   & $\beta$                &                      & 12                    & 2                                   \\
                   & $T_{hi}$ (K)               &                      & 15000                 & 5000                                \\
                   & $T_{lo}$ (K)               &                      & 8000                  & 4000                                \\
MG   J0414$+$0534  & $\chi^2_{min}$         & 20.1                 & 20.0                  & 15.1 ($f=0.09$, $\kappa_\lambda=0.6299$)                     \\
                   & $f_{\rm Edd}$          & 0.07                 & 0.08                  & 0.03                                \\
                   & $L_{bol,opt}$          & 1.368                & 1.484                 & 1.036                               \\
                   & $\beta$                &                      & 1                     & 3                                   \\
                   & $T_{hi}$               &                      & 10000                 & 60000                               \\
                   & $T_{lo}$               &                      & 10000                 & 28000                               \\
HE    0435$-$1223  & $\chi^2_{min}$         & 30.7                 & 29.1                  & 0.9 ($f=0.25$, $\kappa_\lambda=0.4841$)                      \\
                   & $f_{\rm Edd}$          & 0.7                  & 0.8                   & 0.4                                 \\
                   & $L_{bol,opt}$          & 3.972                & 4.532                 & 0.423                               \\
                   & $\beta$                &                      & 6                     & 18                                  \\
                   & $T_{hi}$               &                      & 10000                 & 30000                               \\
                   & $T_{lo}$               &                      & 6000                  & 16000                               \\
RX   J0911$+$0551  & $\chi^2_{min}$         & 135.1                & 128.1                 & 5.4 ($f=0.25$, $\kappa_\lambda=0.3034$)                      \\
                   & $f_{\rm Edd}$          & 1                    & 1                     & 0.4                                 \\
                   & $L_{bol,opt}$          & 9.069                & 8.936                 & 0.883                               \\
                   & $\beta$                &                      & 12                    & 18                                  \\
                   & $T_{hi}$               &                      & 20000                 & 35000                               \\
                   & $T_{lo}$               &                      & 10000                 & 14000                               \\
SDSS J0924$+$0219  & $\chi^2_{min}$         & 37.4                 & 35.5                  & 1.7 ($f=0.16$, $\kappa_\lambda=0.3233$)                      \\
                   & $f_{\rm Edd}$          & 1                    & 1                     & 0.3                                 \\
                   & $L_{bol,opt}$          & 1.263                & 1.261                 & 0.035                               \\
                   & $\beta$                &                      & 6                     & 12                                  \\
                   & $T_{hi}$               &                      & 10000                 & 25000                               \\
                   & $T_{lo}$               &                      & 6000                  & 12000                               \\
HE    1113$-$0641  & $\chi^2_{min}$         & 13.2                 & 12.8                  & 2.4 ($f=0.36$, $\kappa_\lambda=0.6448$)                      \\
                   & $f_{\rm Edd}$          & 0.7                  & 0.8                   & 1                                   \\
                   & $L_{bol,opt}$          & 0.699                & 0.798                 & 0.197                               \\
                   & $\beta$                &                      & 6                     & 12                                  \\
                   & $T_{hi}$               &                      & 10000                 & 35000                               \\
                   & $T_{lo}$               &                      & 6000                  & 16000                               \\
PG    1115$+$080   & $\chi^2_{min}$         & 62.3                 & 65.8                  & 2.7 ($f=0.04$, $\kappa_\lambda=0.1753$)                      \\
                   & $f_{\rm Edd}$          & 1                    & 1                     & 1                                   \\
                   & $L_{bol,opt}$          & 13.892               & 13.718                & 0.956                               \\
                   & $\beta$                &                      & 3                     & 18                                  \\
                   & $T_{hi}$               &                      & 15000                 & 15000                               \\
                   & $T_{lo}$               &                      & 10000                 & 12000                               \\
RX   J1131$-$1231  & $\chi^2_{min}$         & 34.2                 & 23.7                  & 14.6 ($f=0.36$, $\kappa_\lambda=0.7578$)                     \\
                   & $f_{\rm Edd}$          & 0.3                  & 0.4                   & 0.2                                 \\
                   & $L_{bol,opt}$          & 0.206                & 0.275                 & 0.032                               \\
                   & $\beta$                &                      & 6                     & 12                                  \\
                   & $T_{hi}$               &                      & 10000                 & 35000                               \\
                   & $T_{lo}$               &                      & 6000                  & 16000                               \\
SDSS J1138$+$0314  & $\chi^2_{min}$         & 53.8                 & 53.5                  & 29.7 ($f=0.16$, $\kappa_\lambda=0.3073$)                     \\
                   & $f_{\rm Edd}$          & 1                    & 1                     & 1                                   \\
                   & $L_{bol,opt}$          & 0.461                & 0.460                 & 0.038                               \\
                   & $\beta$                &                      & 6                     & 12                                  \\
                   & $T_{hi}$               &                      & 10000                 & 35000                               \\
                   & $T_{lo}$               &                      & 6000                  & 16000                               \\
WFI  J2026$-$4536  & $\chi^2_{min}$         & 78.8                 & 75.8                  & 16.3 ($f=0.25$, $\kappa_\lambda=0.2972$)                     \\
                   & $f_{\rm Edd}$          & 1                    & 1                     & 1                                   \\
                   & $L_{bol,opt}$          & 8.956                & 8.851                 & 1.327                               \\
                   & $\beta$                &                      & 12                    & 12                                  \\
                   & $T_{hi}$               &                      & 15000                 & 25000                               \\
                   & $T_{lo}$               &                      & 8000                  & 12000                               \\
WFI  J2033$-$4723  & $\chi^2_{min}$         & 299.4                & 300.0                 & 111.0 ($f=0.04$, $\kappa_\lambda=0.1047$)                    \\
                   & $f_{\rm Edd}$          & 1                    & 1                     & 1                                   \\
                   & $L_{bol,opt}$          & 2.062                & 2.058                 & 0.063                               \\
                   & $\beta$                &                      & 6                     & 6                                   \\
                   & $T_{hi}$               &                      & 10000                 & 15000                               \\
                   & $T_{lo}$               &                      & 6000                  & 8000                                \\
\hline
\end{tabular}\\
See Appendix \ref{sizes} for a description of $\kappa_\lambda$, and
\S~\ref{results} for details on other quantities presented herein.
\label{tab:chi2min}
\end{minipage}
\end{table*}

\subsection{Phenomenological model results} \label{results}

For each quasar we calculate a temperature profile for all allowable 
combinations of the specified parameter values given for each model above.
For each model --- SS73, forming disc, and magnetically restrained
for values of $f$ from 0 to 0.81 --- we find the minimum $\chi^2$ value
over all tested values of \fedd, $T_{hi}$, $T_{lo}$ and $\beta$.
We plot the resulting half-light radii vs. wavelength, SEDs, and
$\chi^2_{min}$ values for each quasar individually in Figure \ref{fig:newer};
see the figure caption for details.
We present the $\chi^2_{min}$ values and best-fit parameter values for 
the SS73, forming disc, and best-fit magnetically restrained models
in Table \ref{tab:chi2min}.
For ease of comparison with other size values, Table \ref{tab:2} presents
$I$-band flux sizes $r_I$ for both the $\beta=3/4$ case and the best-fit 
magnetically restrained models, calculated following Appendix~\ref{sizes}.

The first point to notice about the best fits is that
the best-fit SS73 and forming disc models usually
have large \fedd\ (0.7 $\leq$ \fedd $\leq$ 1 in nine out of eleven cases).
Larger \fedd\ results in a larger accretion disc, making it easier to match
the large inferred half-light radii.
This effect is less pronounced in the best-fit magnetically restrained models;
they have \fedd=1 in six of eleven cases, but three of those are poor fits.
(However, we have only considered face-on discs in our modelling.
An inclined disc will have half-light radii smaller than it does when face-on,
and given that quasar discs will have a nonzero average inclination angle,
the true best-fit \fedd\ of our objects in all models will, statistically, be
somewhat larger than the values quoted.)

The second point to notice 
is that the forming disc model is 
never a significantly better fit than the SS73 model.

The third point to notice is that in eight of eleven cases,
the magnetically restrained models can provide a fit
which is better than the SS73 or forming disc fit
and is also a statistically acceptable fit at the 99.73\% confidence level
if the full errors $\sigma$ from B11 are used in the $\chi^2$ calculations.
(If the $\sigma_r$ values are used instead,
only in five of eleven cases are statistically acceptable fits found.)

Two of the cases without acceptable fits using $\sigma$, HE 0230 and WFI J2033, 
have unusual wavelength dependences of their half-light radii and are 
discussed below.
The third case, SDSS J1138, has the lowest $M_{BH}$ 
and the largest ratio of $L_{bol,opt}$ to $M_{BH}$ in our sample.
The SS73 and forming disc models can match the luminosity of SDSS J1138,
but not its large half-light radii, even with \fedd=1.  
The magnetically restrained models with \fedd=1 can better
match the shape of its half-light radius vs. wavelength curve, but 
still cannot match their large values, and is a worse fit to the 
observed luminosity.
The large $\chi^2_{min}$ in this object could be reduced if
future observations reveal that its $M_{BH}$ is larger than the value we adopt,
which is quite plausible given the considerable uncertainties on $M_{BH}$
estimates \citep[see, e.g., the comprehensive review of][]{2013BASI...41...61S}.
The same might be true of HE 0230 and WFI J2033, which 
in addition to the unusual wavelength dependences of their half-light radii 
have the next largest ratios of $L_{bol,opt}$ to $M_{BH}$ in our sample.

However, Table \ref{tab:2} shows that
the flux sizes of the best-fit magnetically restrained models 
are still about an order of magnitude smaller, on average,
than the measured microlensing sizes.
(They are larger than the flux sizes of a $\beta=3/4$ disc
by $\Delta\log r_{1/2}=+0.37$ on average.)
The one exception is HE~0230$-$2130, which has a best-fit
magnetically restrained model with $f=0$, corresponding to 
negligible thermal emission within the radius of the temperature spike.
Except in that case, our best-fit magnetically restrained models 
produce more flux than observed in these objects.
Incorporating flux sizes as a constraint in the fitting might shift
the best-fit $f$ values to $f=0$ in many 
cases, as the $\chi^2$ minimum as a function of $f$ is quite broad.

The best-fit SS73 and forming disc models tend to overestimate $L_{bol,opt}$,
while the best-fit magnetically restrained model tends to slightly
underestimate it.  Better constraints on the models might be obtainable 
by including multiwavelength SEDs for these quasars in the $\chi^2$ fits,
rather than just the total $L_{bol,opt}$.
In the cases of HE 0230, PG 1115 and WFI J2033,
such constraints would considerably increase the $\chi^2$
of the best-fit magnetically restrained models; those best-fit models 
have SEDs that peak closer to 10$^4$\,\AA\ than 10$^3$\,\AA.
It is not clear if a magnetically restrained model would still yield a low
$\chi^2_{min}$ for PG 1115 if measurements of its multiwavelength SED were 
available for consideration in the fitting.

There are some similarities among
the eight acceptable magnetically restrained model fits.
Those best fits have $0.04\leq f\leq 0.36$, corresponding to
temperatures betweeen 20\% and 60\% of the SS73 value.
The best-fit $\beta$ is 12 or 18, except for MG~J0414 ($\beta=3$).
(Note that MG~J0414 is also consistent with an SS73 or forming disc.)
The best-fit $T_{lo}$ is between 12000~K and 16000~K, 
again except for MG~J0414 ($T_{lo}$=28000~K).
The best-fit $T_{hi}$ is between 25000~K and 35000~K, 
except for MG~J0414 ($T_{hi}$=60000~K)
and PG~1115 ($T_{hi}$=15000~K).
The best-fit values of \fedd, however, range from 0.03 to 1.

We can therefore characterize the best-fit temperature profile in
seven of eleven cases as a magnetically restrained model with
a temperature (40$\pm$20)\% of the SS73 value 
within the radius where the SS73 disc reaches (14000$\pm$2000)~K,
at which radius it has a temperature spike reaching (30000$\pm$5000)~K,
which falls off as a power law $\propto R^{-(15\pm 3)}$.

\bigskip

\subsubsection{Quasars with half-light radii which decrease with increasing wavelength} \label{astroid}

Two quasars in B11 (HE~0230 and WFI~J2033) appear to
have half-light radii that decrease with increasing wavelength,
albeit with large uncertainties.
A temperature profile which increases with radius and then abruptly decreases
can in principle reproduce the sign of this trend.  
(Conceptually, such a temperature profile might arise if the accretion
luminosity is dominated by a shock at some radius $R_b$, 
with cooling gas undergoing radial infall within $R_b$.)

The two quasars above have maximum $\Delta \log r_{1/2} = -0.61\pm 0.54$ 
and $\Delta \log r_{1/2} = -0.45\pm 0.54$, respectively,
from $\lambda_{min}$ to $\lambda_{max}$.
However, a radially inverted temperature profile like that described above
cannot explain changes in $r_{1/2}$ greater
than about $\Delta \log r_{1/2} \simeq -0.15$.
That $\Delta \log r_{1/2}$ corresponds to the case of 
a narrow temperature spike at $r_{1/2}(\lambda_{min})$ 
which generates all the luminosity at $\lambda_{min}$,
and a flat temperature profile within that radius
which dominates the luminosity observed at $\lambda_{max}$,
yielding $r_{1/2}(\lambda_{max})=\frac{1}{\sqrt{2}}r_{1/2}(\lambda_{min})$,
so that $\Delta \log r_{1/2} = -\log\sqrt{2} \simeq -0.15$.
A temperature profile increasing with radius within $r_{1/2}(\lambda_{min})$ 
will produce an even smaller $\Delta \log r_{1/2}$ value.
A temperature profile decreasing with radius within $r_{1/2}(\lambda_{min})$
before spiking to a high temperature at that radius
will not match the observed smooth change in $r_{1/2}$ with wavelength;
such a temperature profile would yield $r_{1/2}$ values
decreasing and then increasing with wavelength.

B11 note that the anomalous wavelength dependence of these two objects'
size estimates might also be produced by unusual microlensing caustic patterns
such as those found at the center of an astroid caustic.
(The magnification increases away from the center of an astroid caustic,
leading to larger flux anomalies for larger sources.)
That explanation for these objects' half-light radii
seems more likely after consideration of the difficulty in
reproducing the results with actual disc temperature profiles.
We predict that further observations of these two objects will not show the 
anomalous size estimates, as long as the time baseline between old and new
observations is sufficient for the unusual microlensing caustic pattern to 
have moved out of our line of sight to the continuum source.

\begin{figure*}
\includegraphics[angle=0, width=0.440\textwidth]{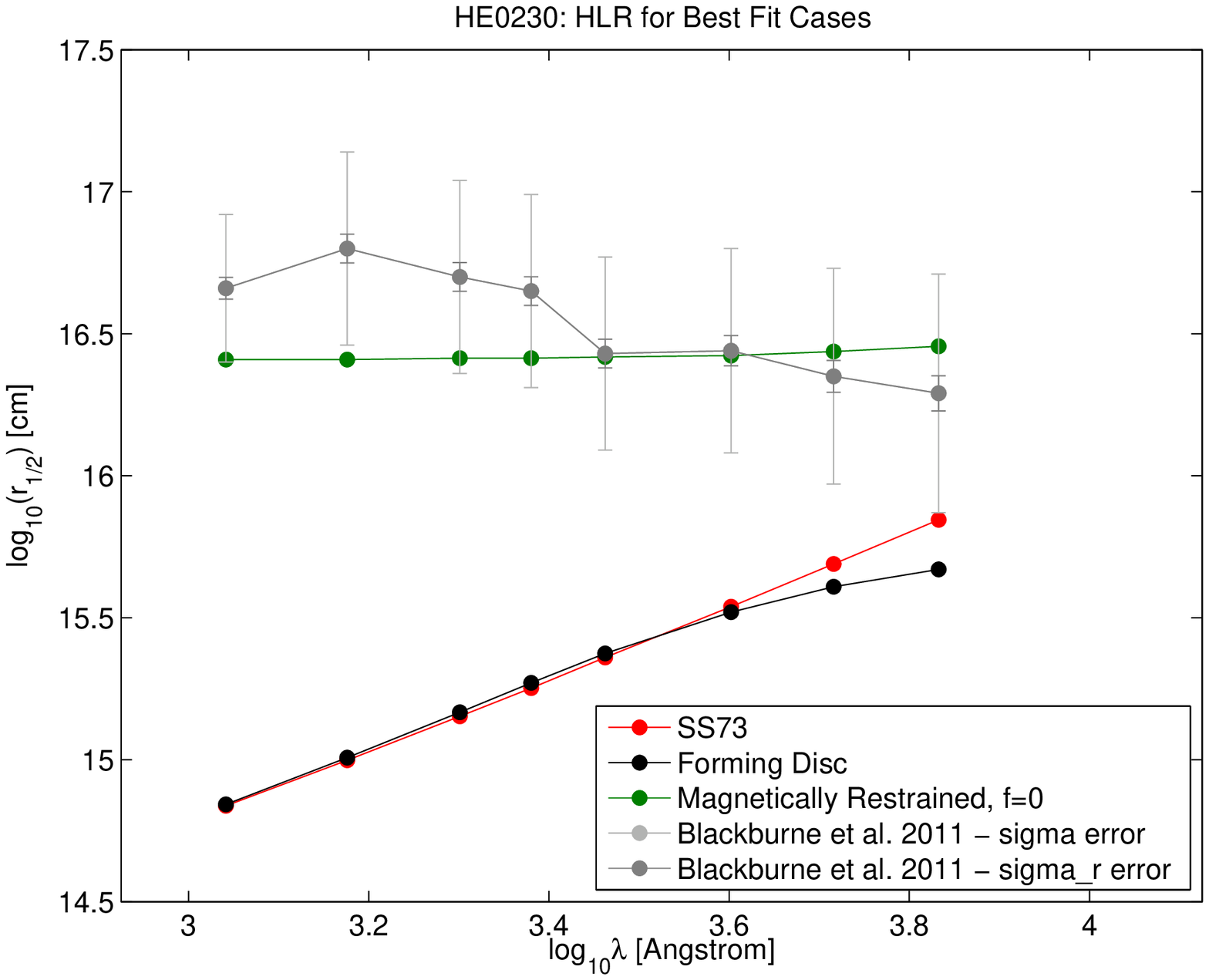}\\
\includegraphics[angle=0, width=0.440\textwidth]{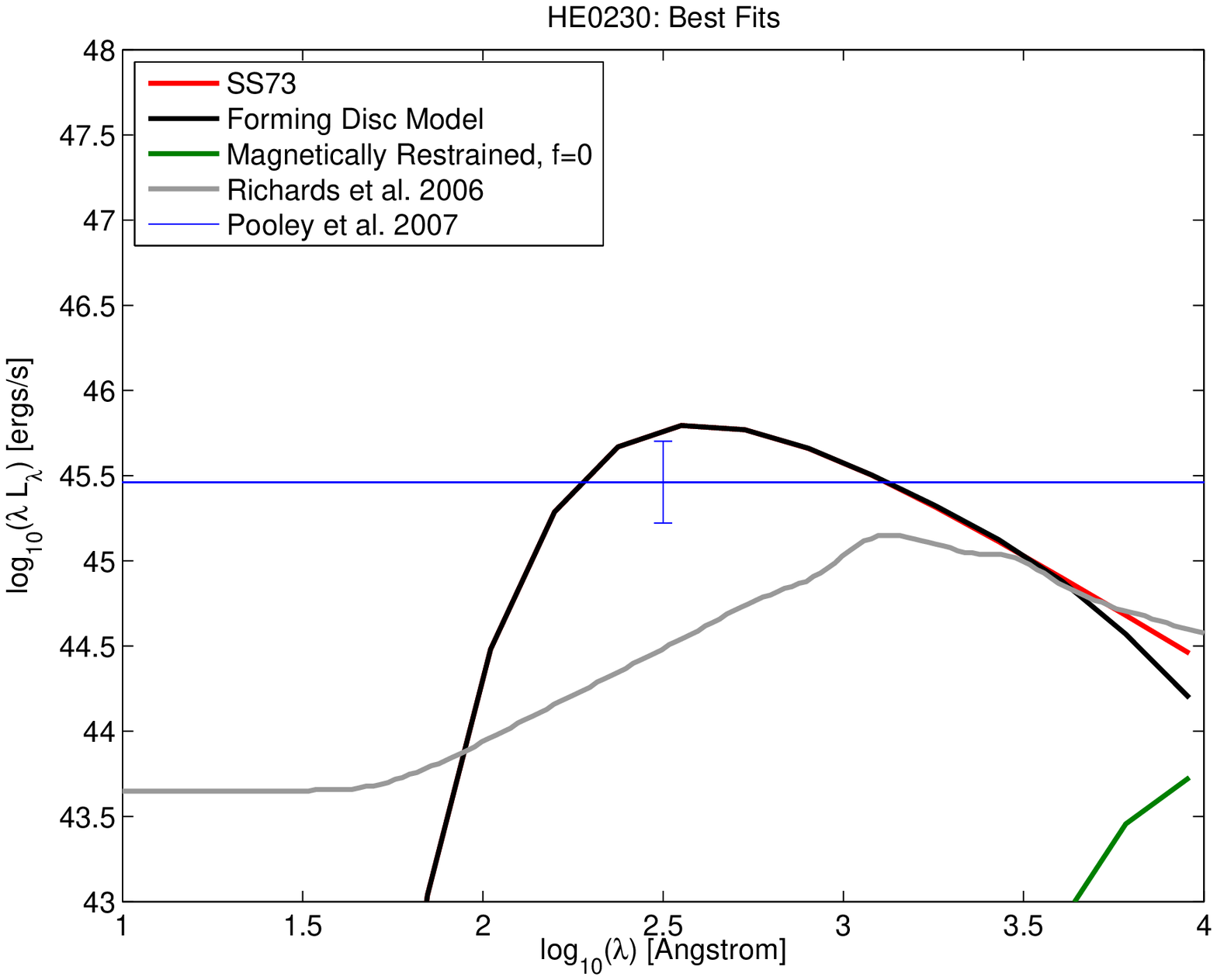}\\
\includegraphics[angle=0, width=0.440\textwidth]{Plot_Chi_vs_F_v3.ps_pages01}
\caption{Half-light radius $r_{1/2}$ vs. wavelength (top row),
spectral energy distribution (middle row),
and $\chi^2_{min}$ results (bottom row)
for each of the quasars studied.
In the $r_{1/2}$ plots, the measured points from B11 are
shown in grey, with the random and full errors on each $r_{1/2}$ value
shown as the shorter and longer error bars, respectively.
The best-fit SS73 model is shown in red,
the best-fit forming disc model in black,
and the best-fit magnetically restrained model in green.
In the SED plots, the quasar's $L_{bol,opt}$ value from Pooley et al. (2007)
is shown as the horizontal blue line.
The mean SED from Richards et al. (2006), summed from 10$^2$\,\AA\ to 
10$^4$\,\AA, is normalized to that value
as a reference for the typical shape of a quasar's ultraviolet/optical SED.
In the $\chi^2_{min}$ plots,
the red open square on the left shows the SS73 disc $\chi^2_{min}$,
the black open square next to it shows the forming disc $\chi^2_{min}$,
and the circles show the $\chi^2_{min}$ for the magnetically restrained disc
model at each value of $f$ tested.
The solid green line segments show, for each model, the upper limit
for a statistically acceptable fit at the 99.73\% (3$\sigma$) confidence level,
using the random errors only.
The dashed green line segments show the same upper limit using
the full errors quoted in B11.
For the magnetically restrained models,
we use black lines to show the 3$\sigma$ confidence limits
for the five free parameters of the model ($\Delta \chi^2 \leq 18.2$):
at 99.73\% confidence, magnetically restrained models underneath the
black lines provide fits as statistically acceptable as the best fit found,
using random errors only (solid black line)
or the full errors from B11 (dashed black line).
Note that we plot confidence limits for all five free parameters, not just
one free parameter, because the best-fit magnetically restrained models as
a function of $f$ typically have differing values of the other parameters.
}
\label{fig:newer} \end{figure*}
\begin{figure*}
\includegraphics[angle=0, width=0.490\textwidth]{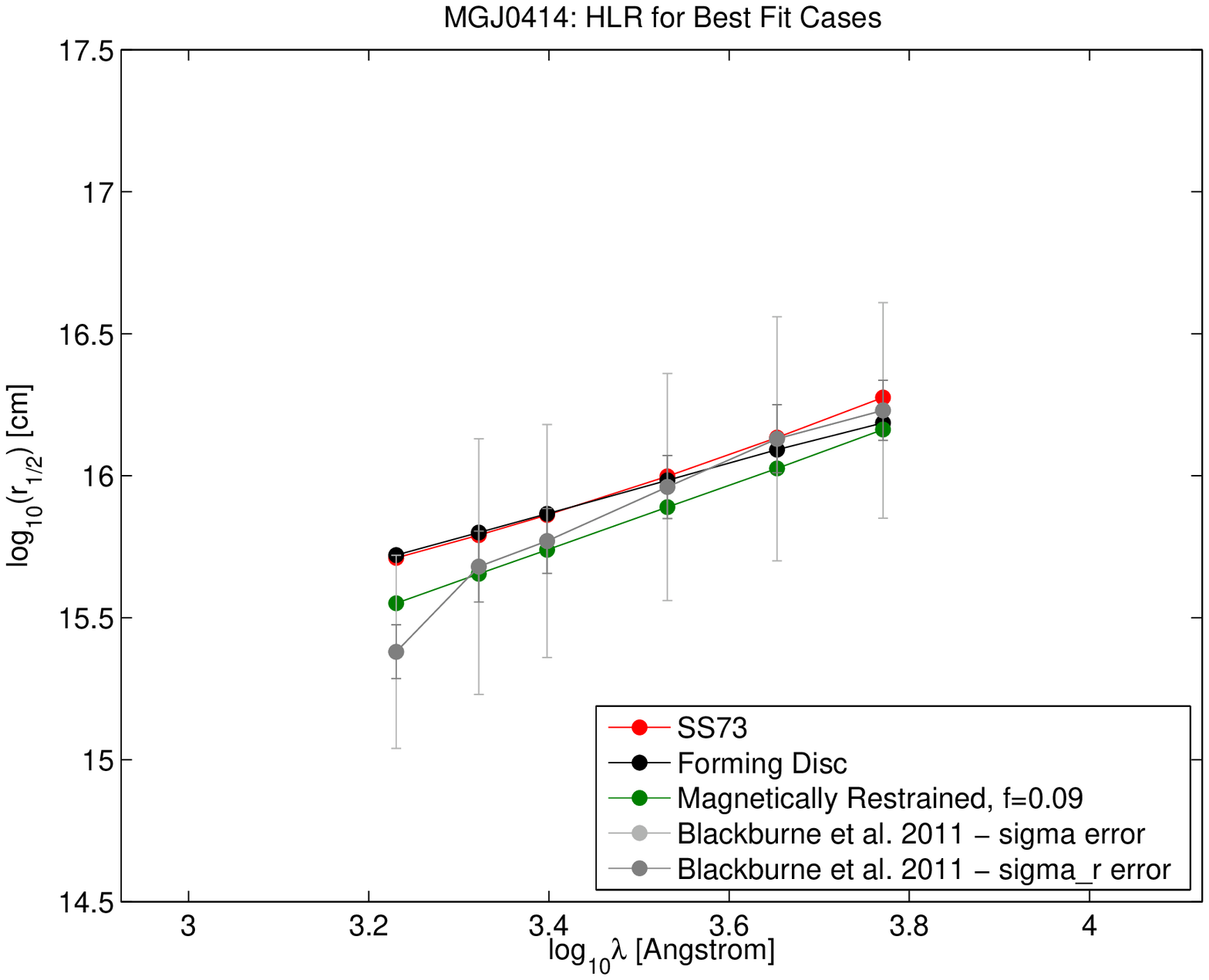}
\includegraphics[angle=0, width=0.490\textwidth]{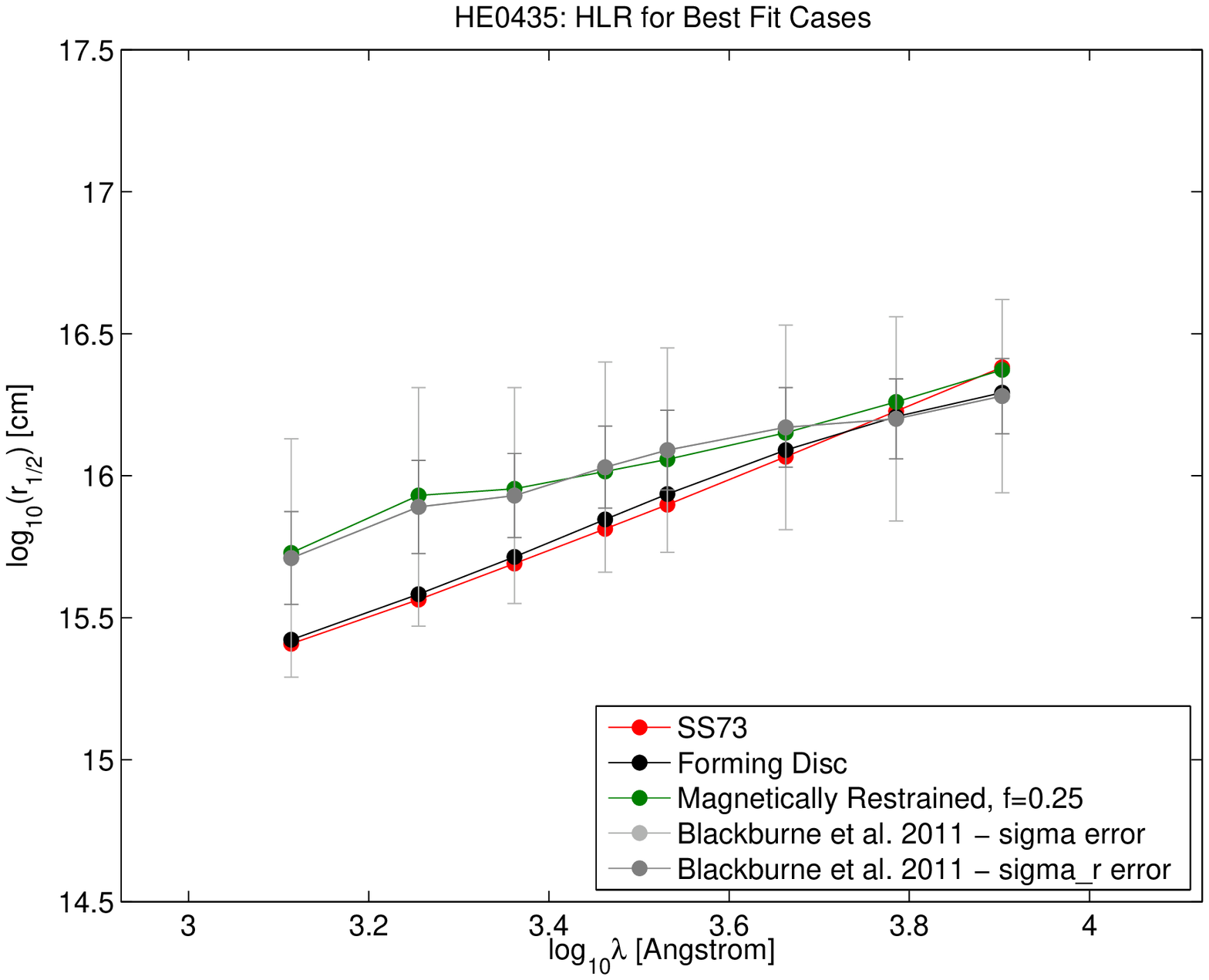}\\
\includegraphics[angle=0, width=0.490\textwidth]{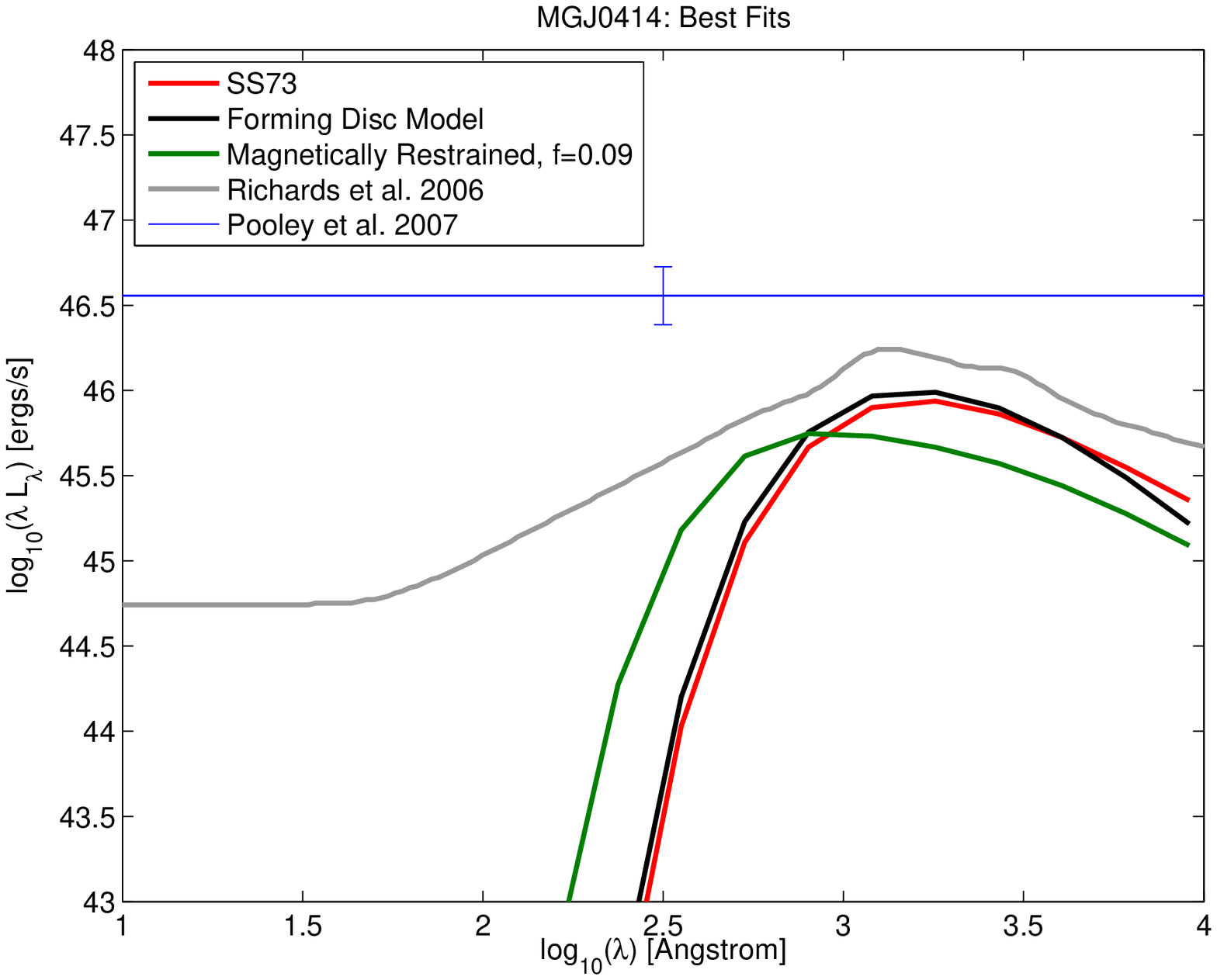}
\includegraphics[angle=0, width=0.490\textwidth]{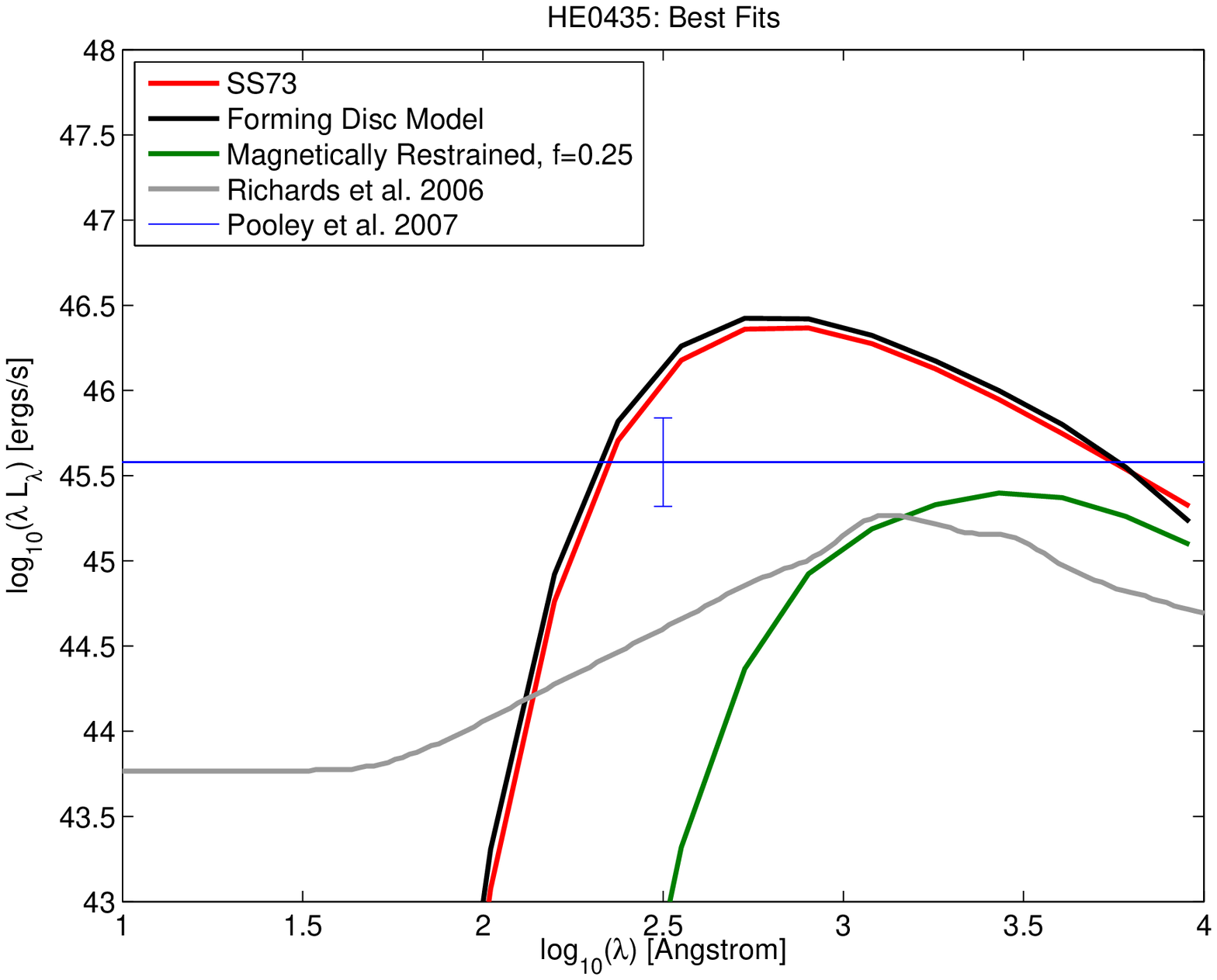}\\
\includegraphics[angle=0, width=0.490\textwidth]{Plot_Chi_vs_F_v3.ps_pages02}
\includegraphics[angle=0, width=0.490\textwidth]{Plot_Chi_vs_F_v3.ps_pages03}
\contcaption{} \end{figure*}
\begin{figure*} 
\includegraphics[angle=0, width=0.490\textwidth]{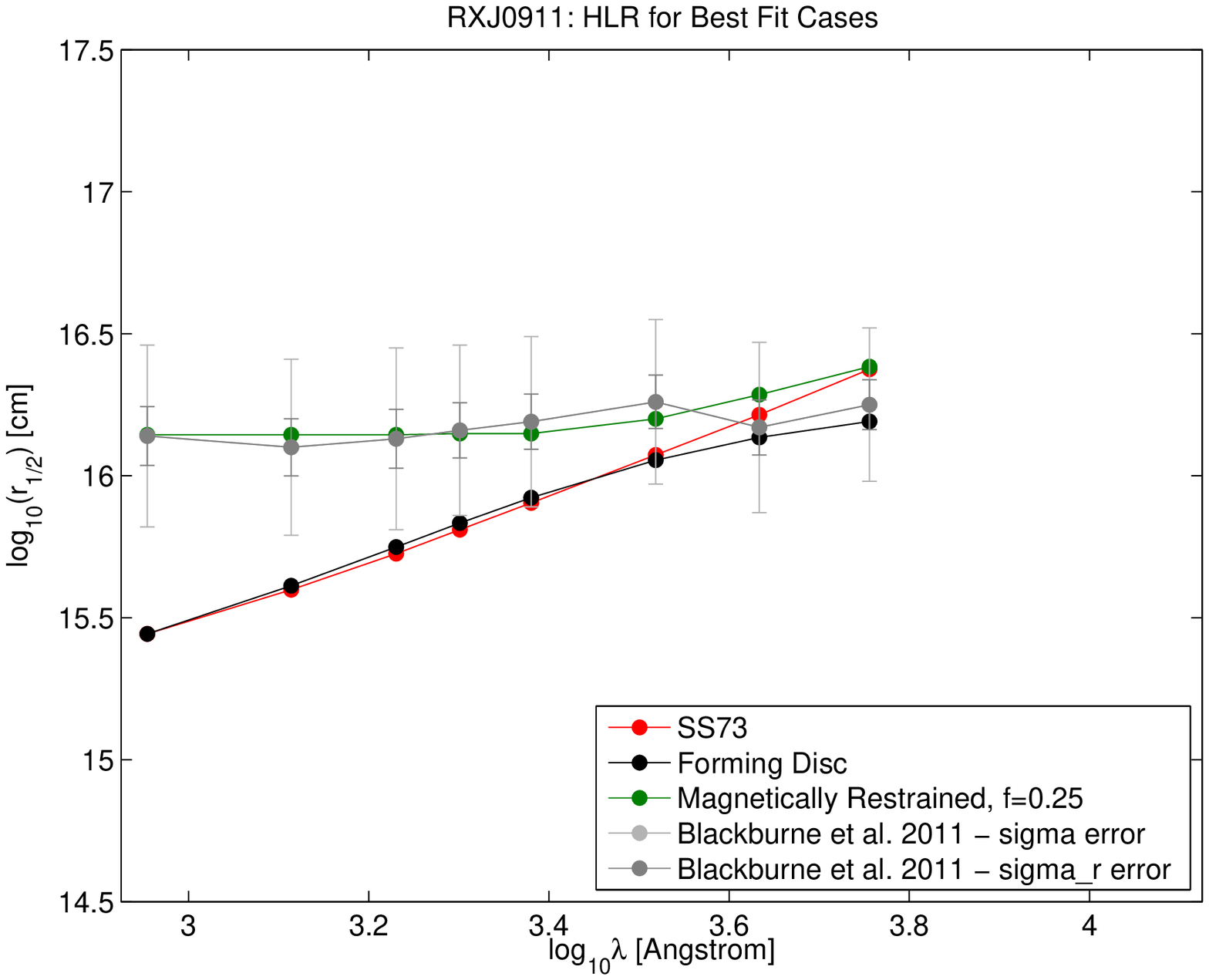}
\includegraphics[angle=0, width=0.490\textwidth]{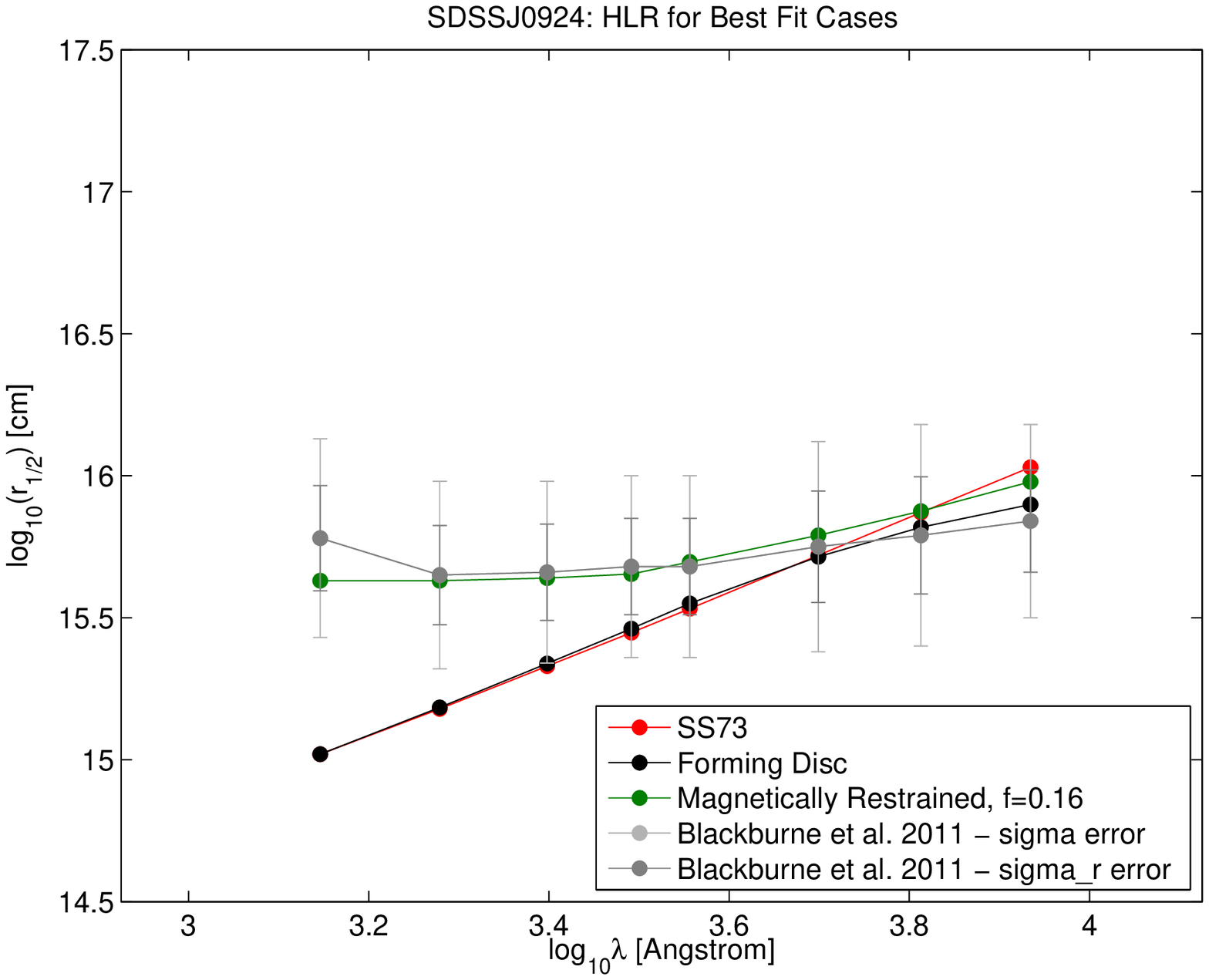}\\
\includegraphics[angle=0, width=0.490\textwidth]{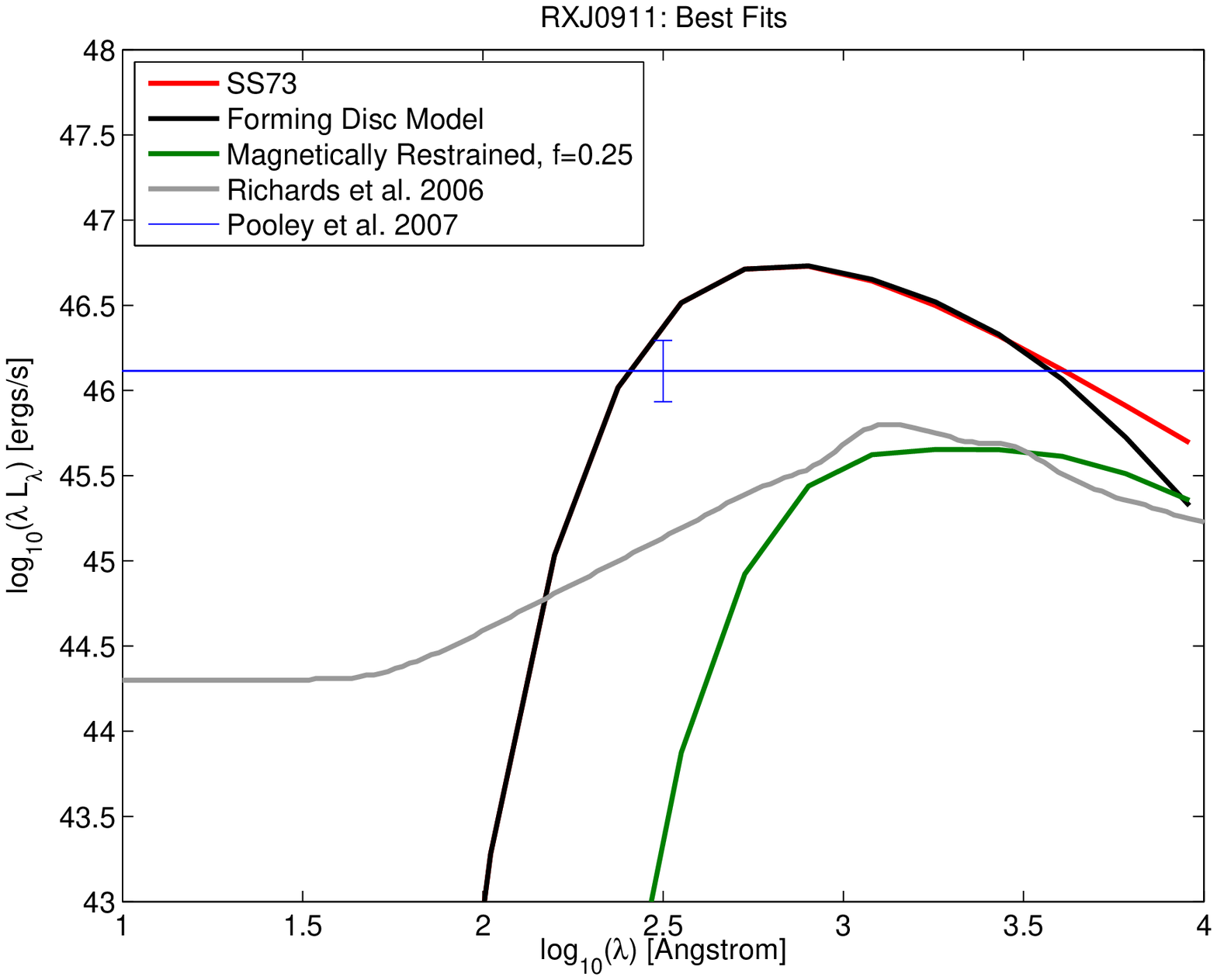}
\includegraphics[angle=0, width=0.490\textwidth]{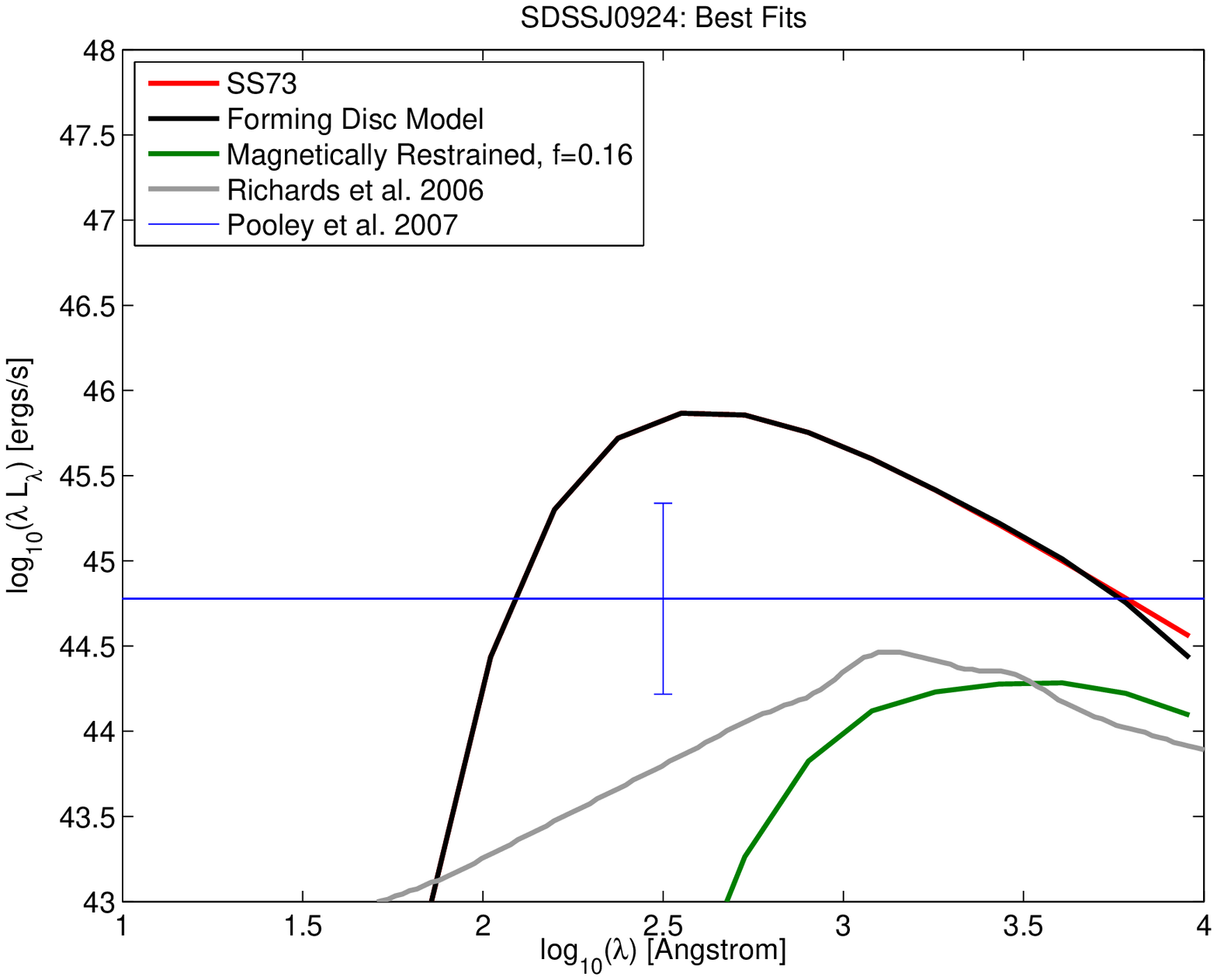}\\
\includegraphics[angle=0, width=0.490\textwidth]{Plot_Chi_vs_F_v3.ps_pages04}
\includegraphics[angle=0, width=0.490\textwidth]{Plot_Chi_vs_F_v3.ps_pages05}
\contcaption{} \end{figure*}
\begin{figure*} 
\includegraphics[angle=0, width=0.490\textwidth]{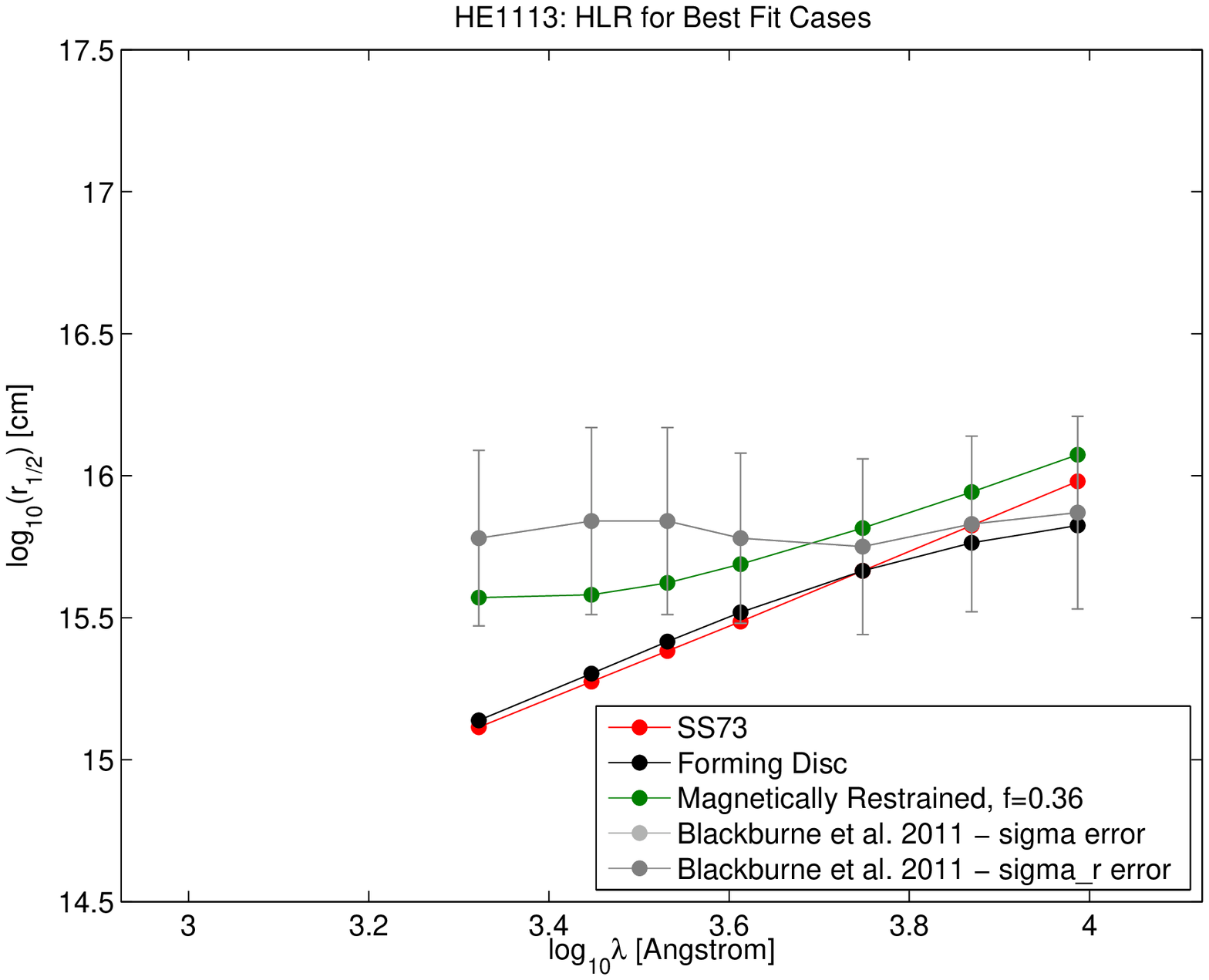}
\includegraphics[angle=0, width=0.490\textwidth]{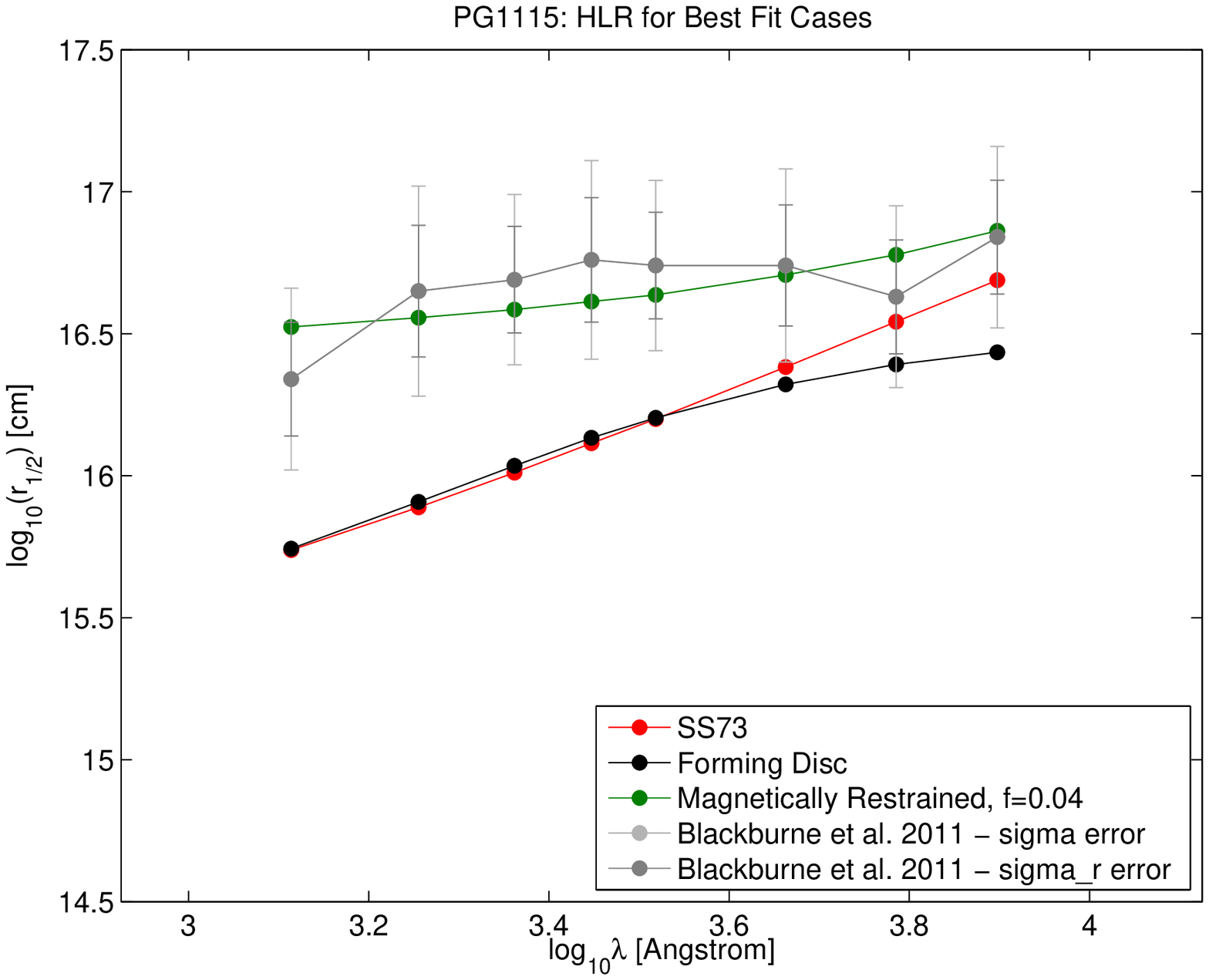}\\
\includegraphics[angle=0, width=0.490\textwidth]{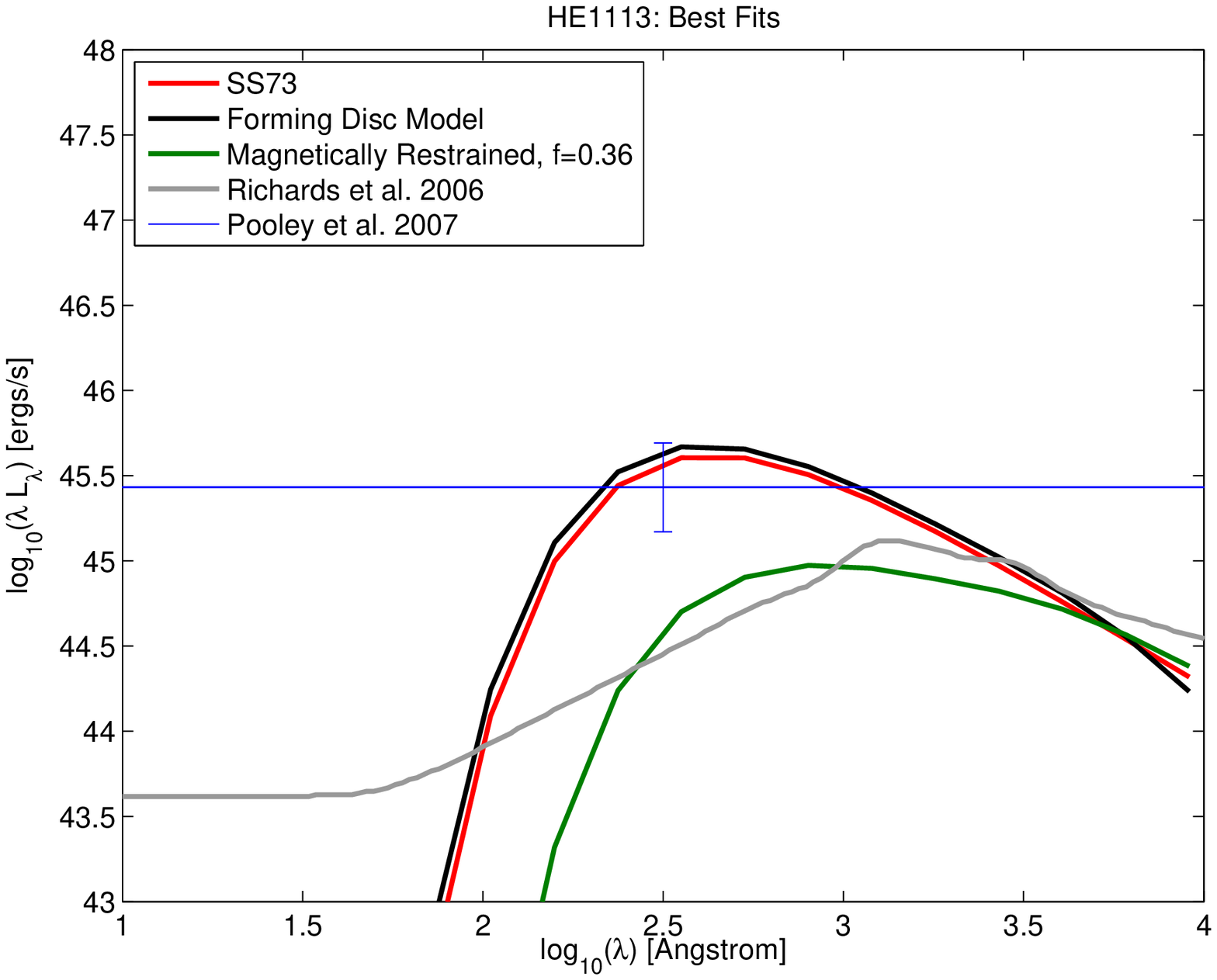}
\includegraphics[angle=0, width=0.490\textwidth]{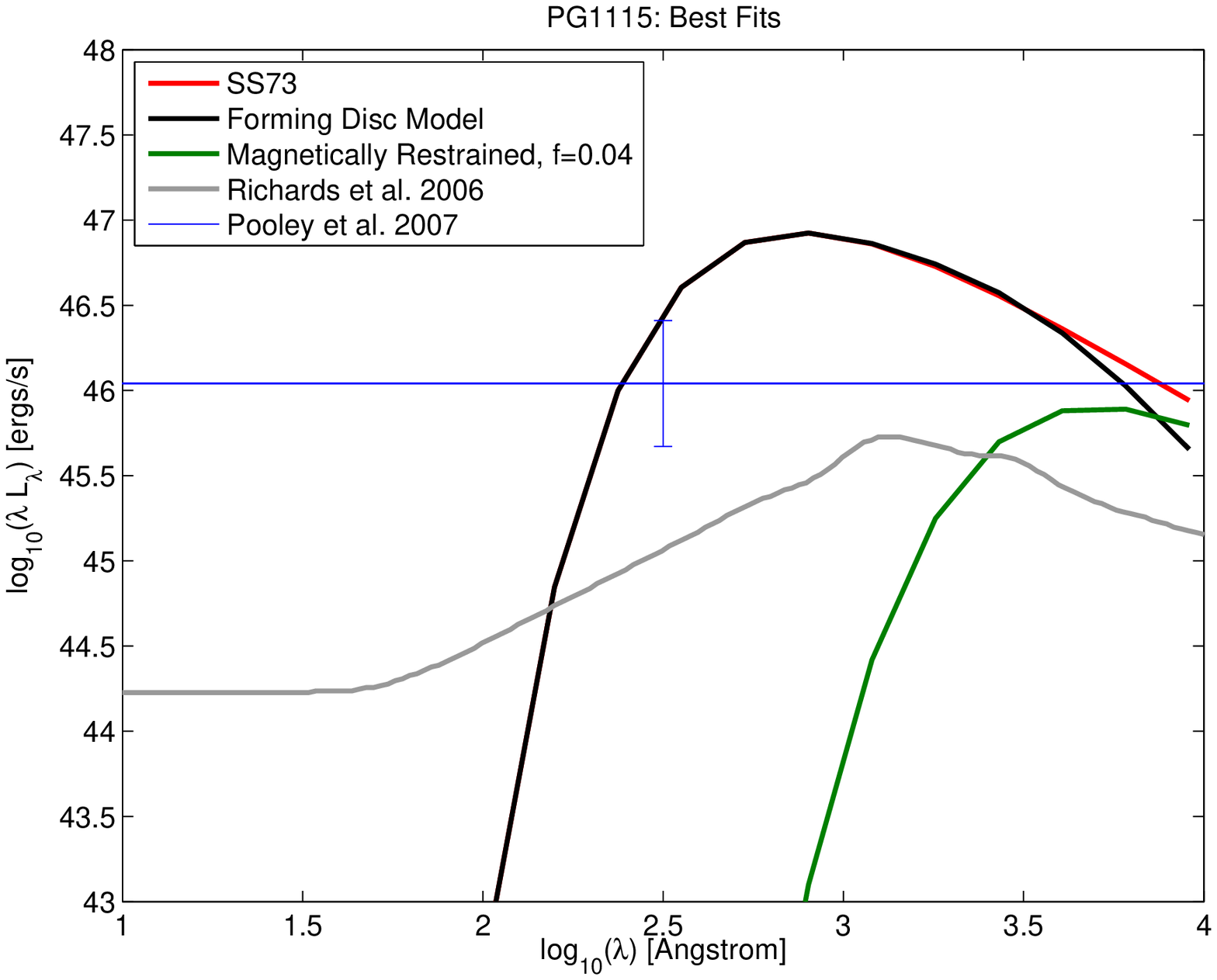}\\
\includegraphics[angle=0, width=0.490\textwidth]{Plot_Chi_vs_F_v3.ps_pages06}
\includegraphics[angle=0, width=0.490\textwidth]{Plot_Chi_vs_F_v3.ps_pages07}
\contcaption{} \end{figure*}
\begin{figure*} 
\includegraphics[angle=0, width=0.490\textwidth]{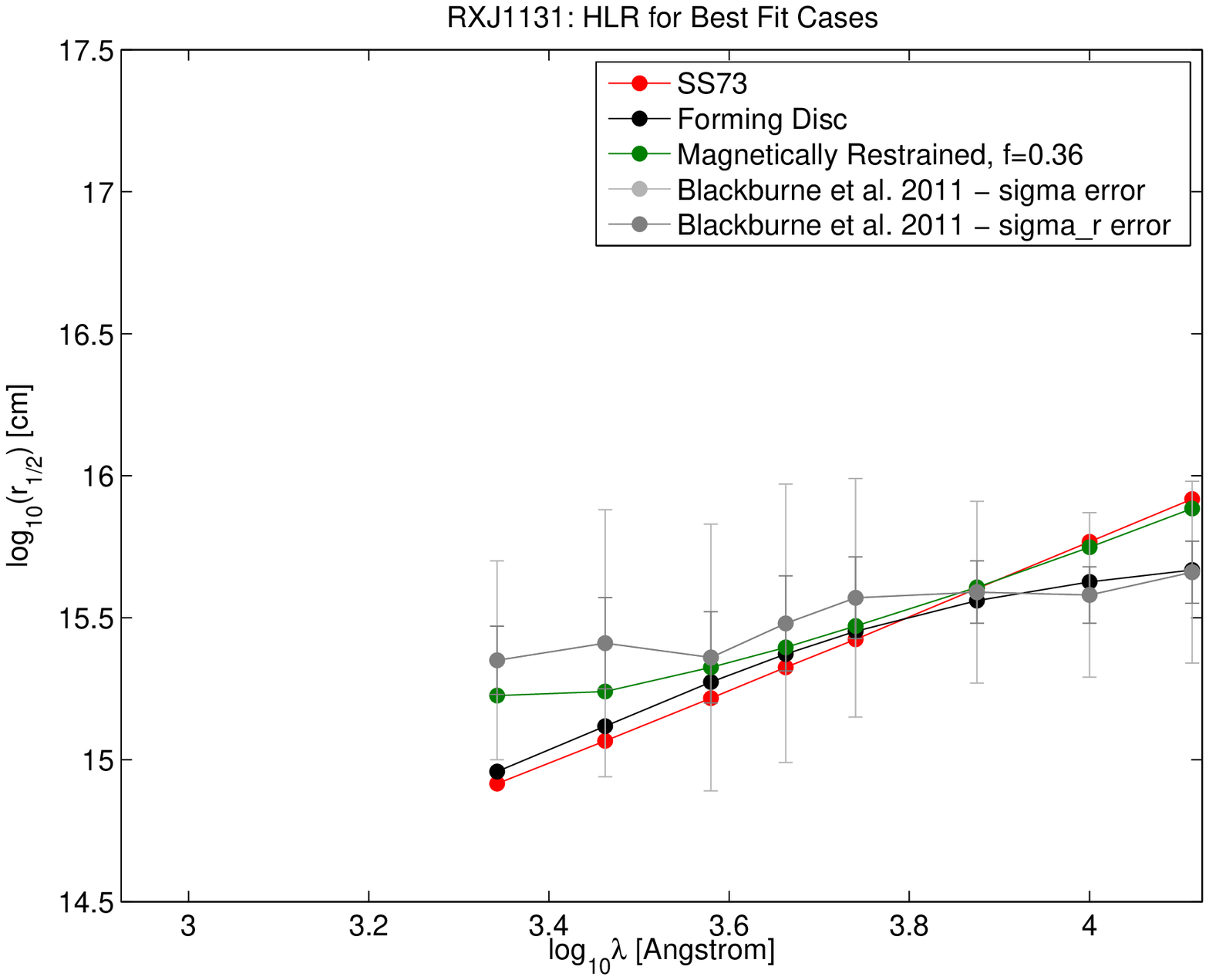}
\includegraphics[angle=0, width=0.490\textwidth]{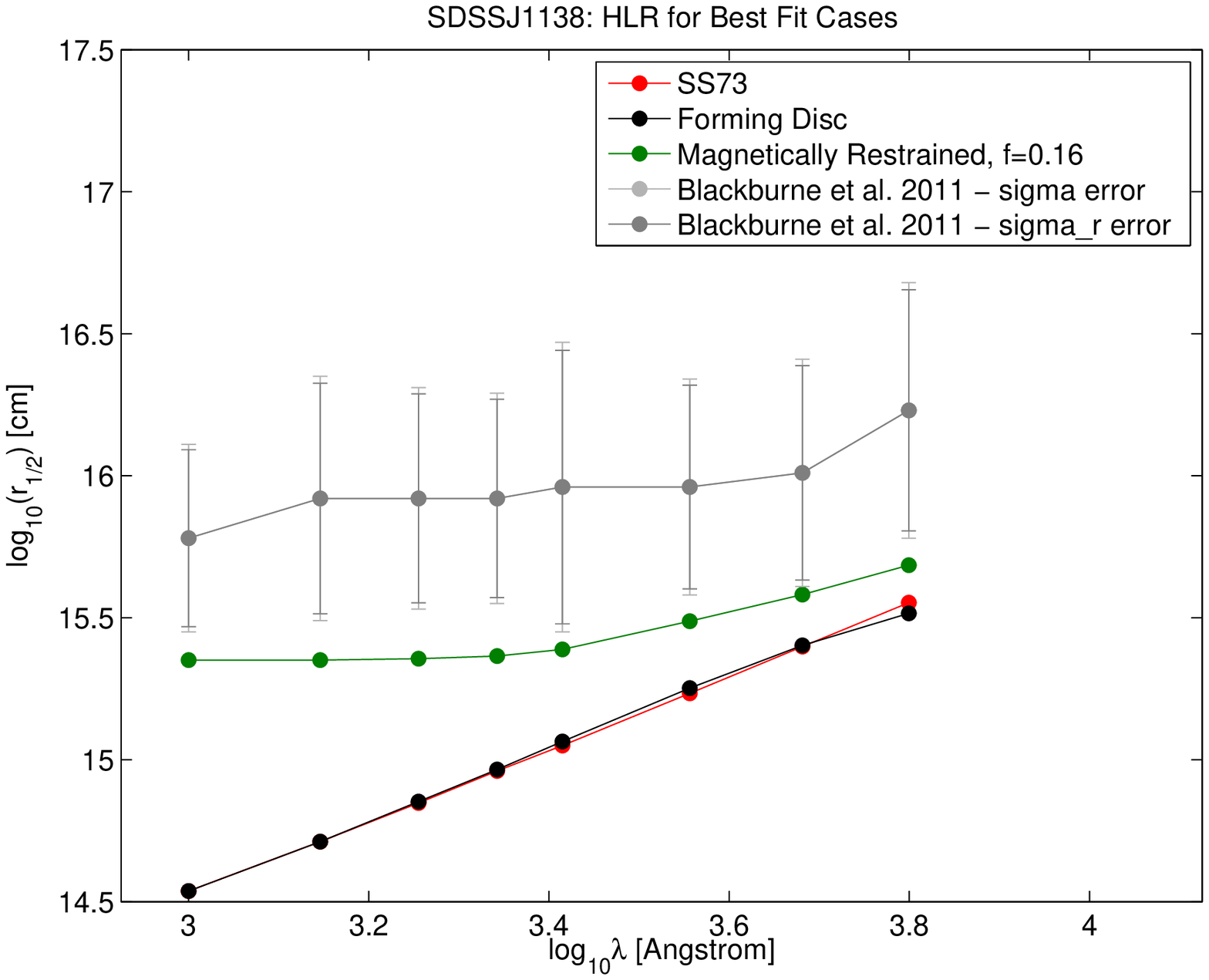}\\
\includegraphics[angle=0, width=0.490\textwidth]{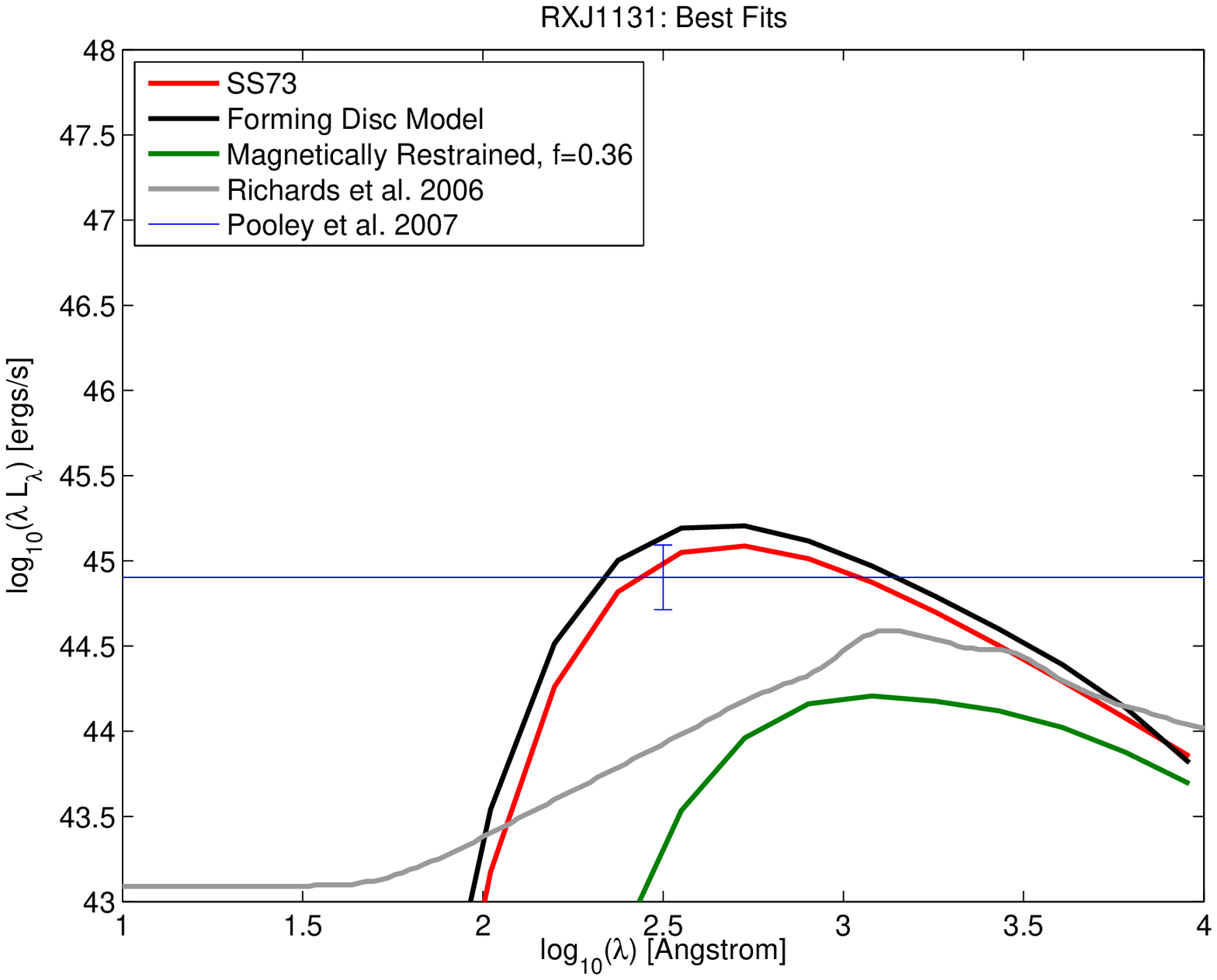}
\includegraphics[angle=0, width=0.490\textwidth]{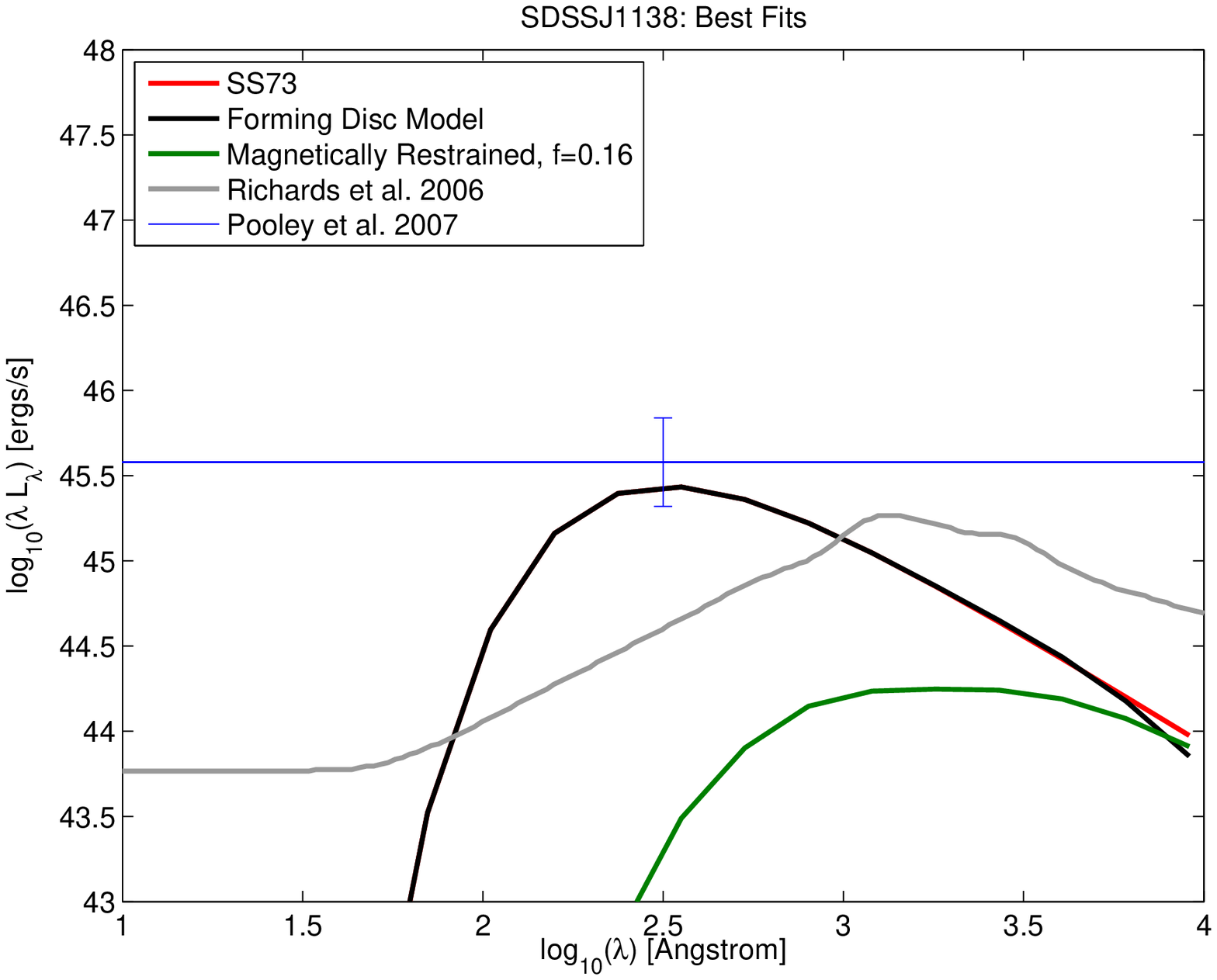}\\
\includegraphics[angle=0, width=0.490\textwidth]{Plot_Chi_vs_F_v3.ps_pages08}
\includegraphics[angle=0, width=0.490\textwidth]{Plot_Chi_vs_F_v3.ps_pages09}
\contcaption{} \end{figure*}
\begin{figure*} 
\includegraphics[angle=0, width=0.490\textwidth]{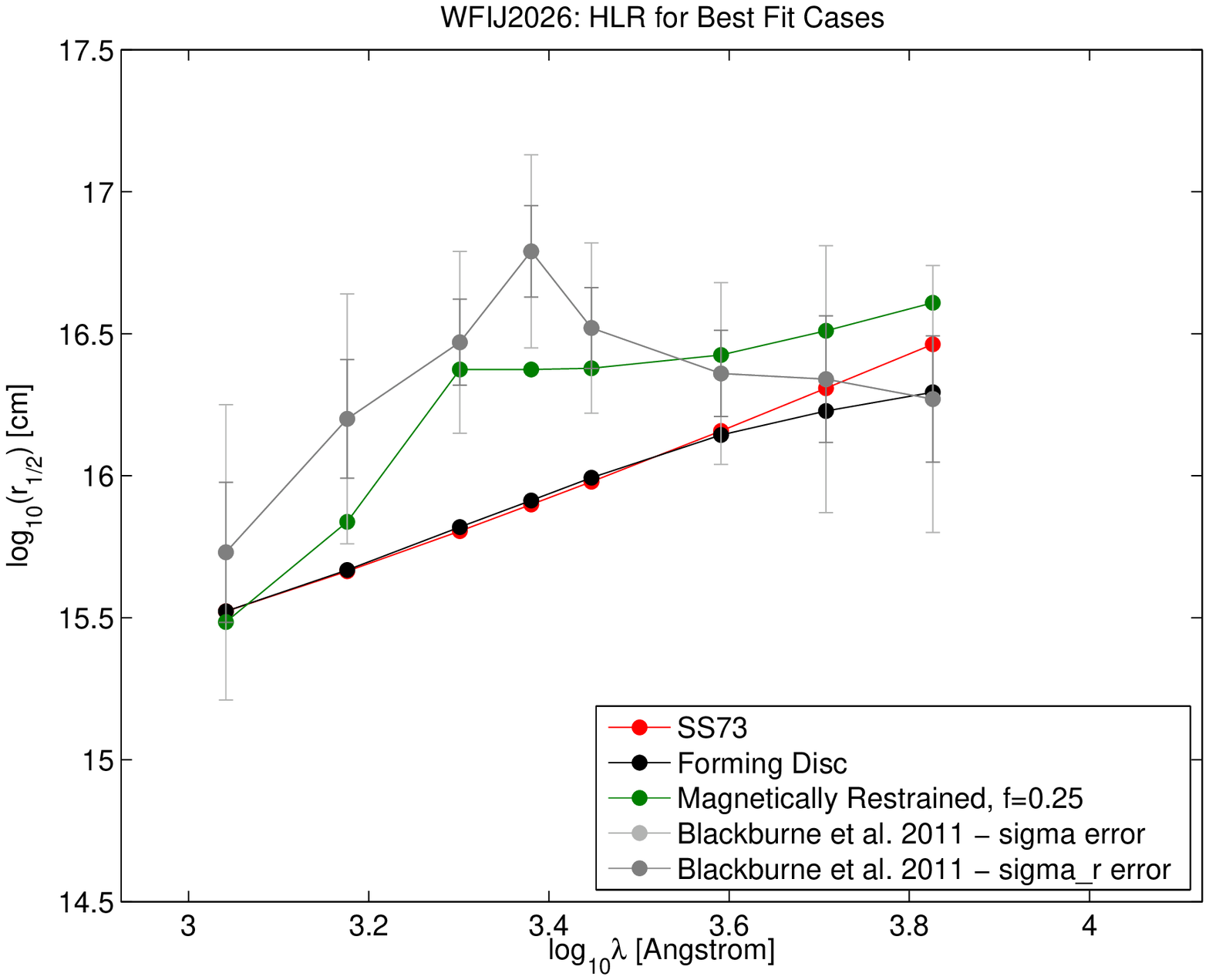}
\includegraphics[angle=0, width=0.490\textwidth]{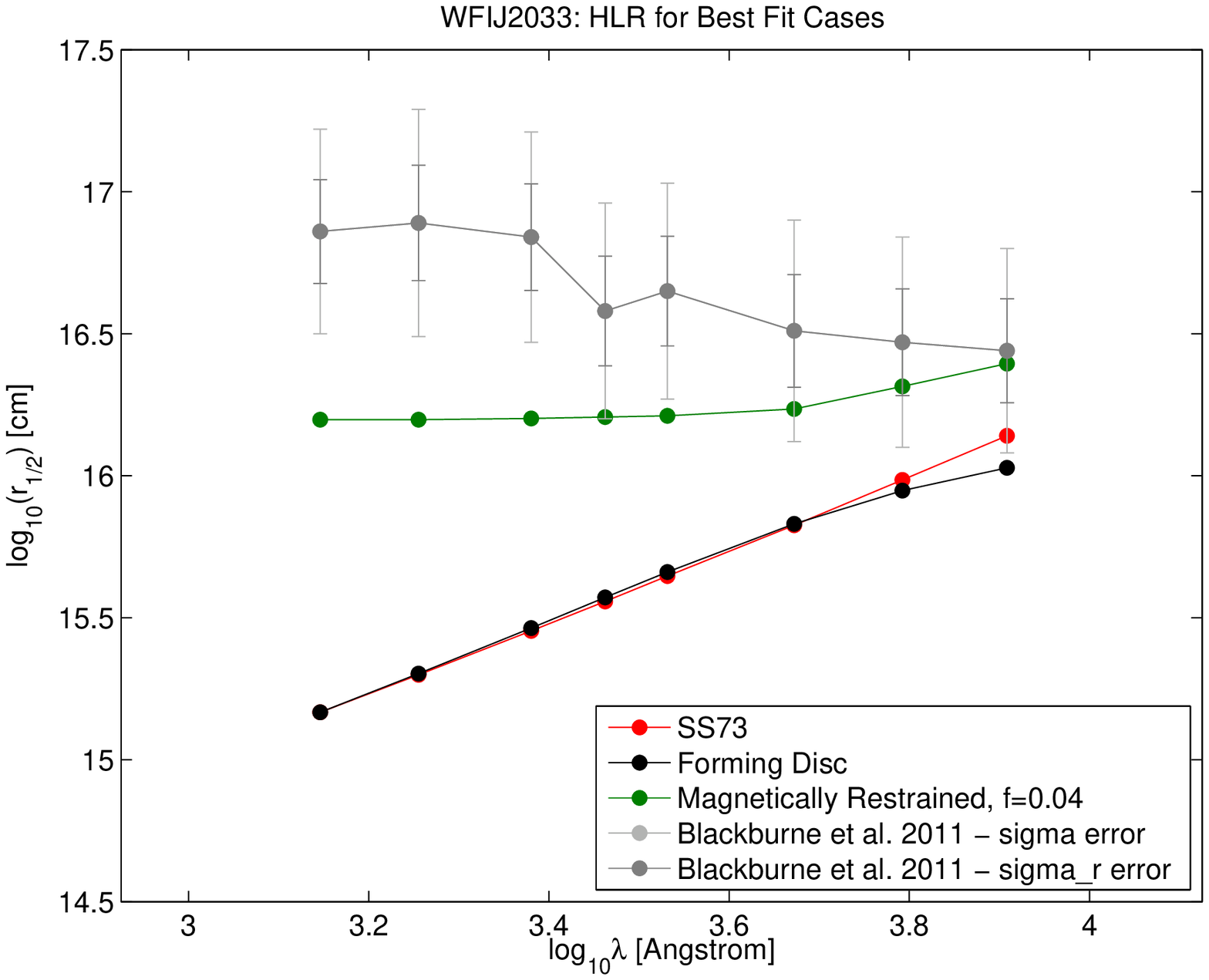}\\
\includegraphics[angle=0, width=0.490\textwidth]{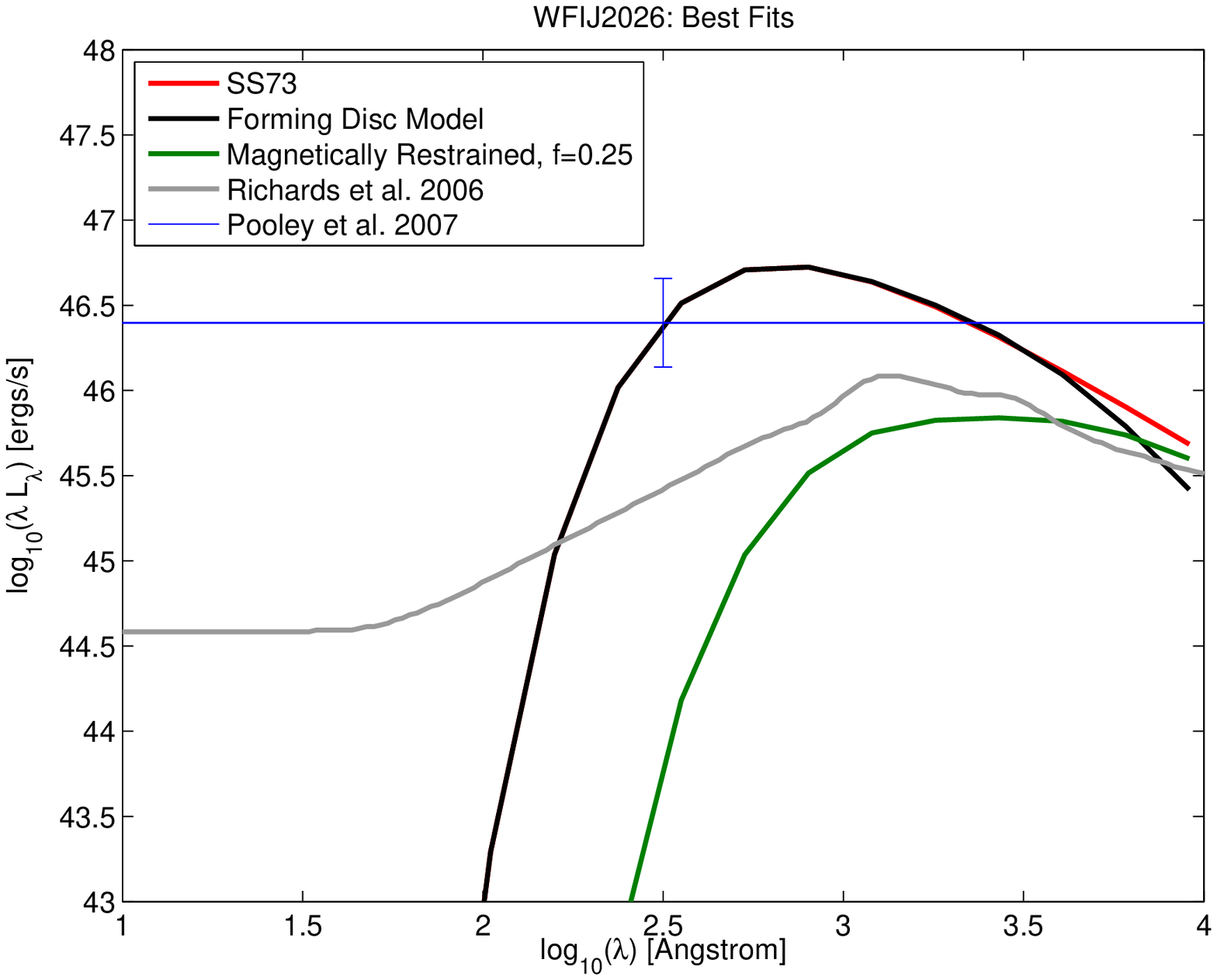}
\includegraphics[angle=0, width=0.490\textwidth]{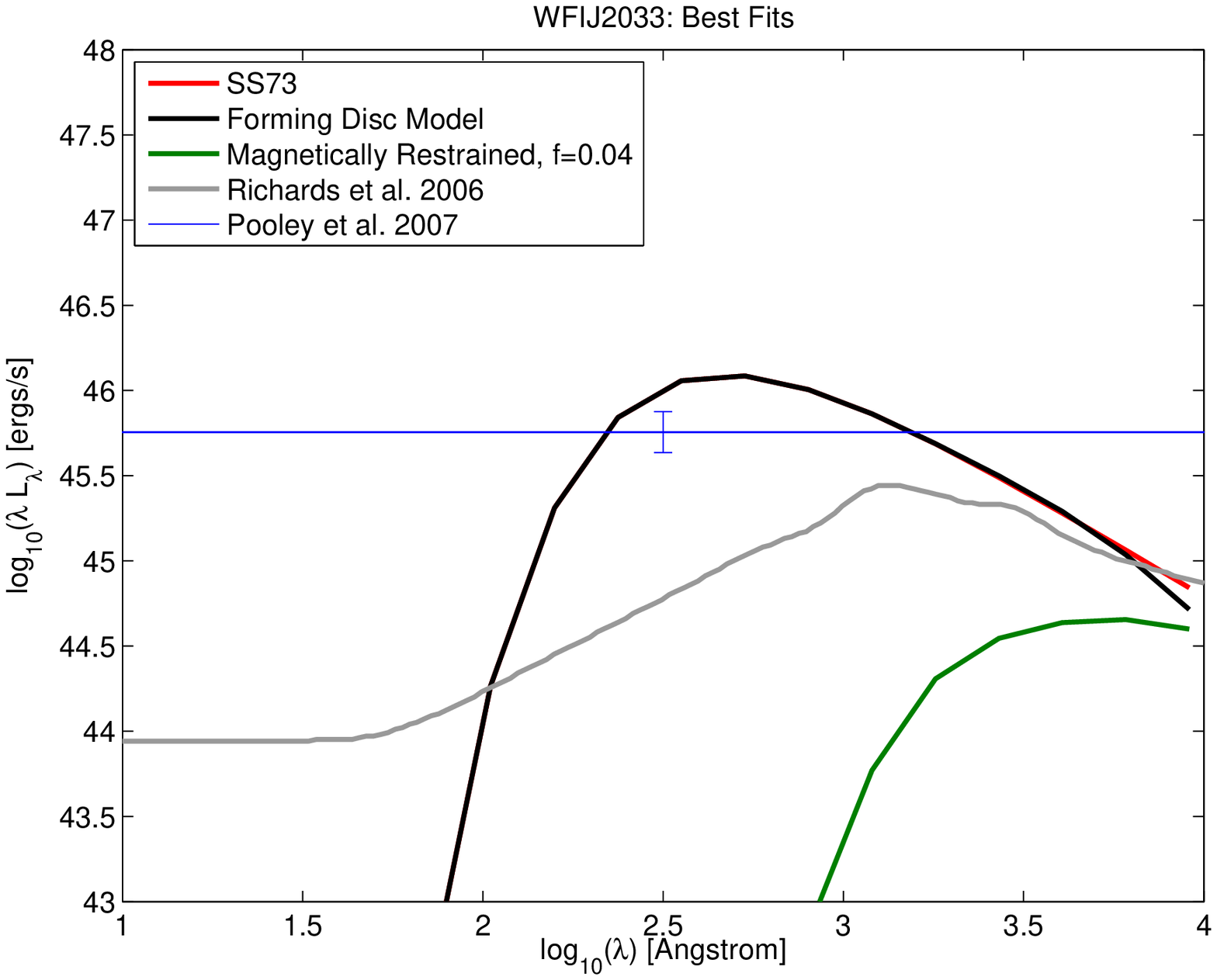}\\
\includegraphics[angle=0, width=0.490\textwidth]{Plot_Chi_vs_F_v3.ps_pages10}
\includegraphics[angle=0, width=0.490\textwidth]{Plot_Chi_vs_F_v3.ps_pages11}
\contcaption{} \end{figure*}

\section{Discussion and conclusions} \label{discuss}

We have tested several energetically plausible quasar accretion disc
temperature profile models against 
the large quasar accretion disc half-light radii
inferred by B11 from microlensing observations.

We have shown that the temperature profiles from our toy model of 
the torn accretion discs proposed by NKPF are not able to 
simultaneously match observed quasar SEDs and the microlensing sizes
of B11.
In our toy model, torn discs yield precessing rings of gas 
orbiting at supersonic speeds which shock when they intersect
and yield SEDs which are much bluer than observed quasar SEDs.

We have demonstrated that acceptable fits (at the 3$\sigma$ level) to the 
half-light radii 
of B11 --- in other words, theory sizes matching the microlensing sizes ---
can be found in eight of eleven cases 
for model discs with a temperature spike at some radius and
lowered temperatures within that radius (our magnetically restrained models).
Acceptable fits can be found in
only four of eleven cases for discs with normal SS73 temperature profiles
and in only three of eleven cases for discs with normal temperature profiles 
inside a temperature spike (our forming disc models).
The poor $\chi^2_{min}$ for both forming disc models 
and magnetically restrained models with large $f$
indicate that the shape of the temperature profile outside the temperature
spike is not a major factor in obtaining an acceptable fit.

We find that the best-fit temperature profile in
seven of eleven cases is a magnetically restrained model with
a temperature (40$\pm$20)\% of the SS73 value
within the radius where the SS73 disc reaches (14000$\pm$2000)~K,
at which radius it has a temperature spike reaching (30000$\pm$5000)~K,
which falls off as a power law $\propto R^{-(15\pm 3)}$.

However, such profiles do not produce flux sizes as large as 
their theory sizes, by about an order of magnitude.
Constraining the magnetically restrained models to explicitly match the
flux sizes may produce better fits by reducing the best-fit $f$ values.

We note that the two quasars in B11 which have half-light radii that
appear to decrease with increasing wavelength cannot be fully explained by
a radially increasing temperature profile with an abrupt cutoff.
They must instead be examples of unusual microlensing patterns;
e.g., astroid caustics (\S~\ref{astroid}).

Future progress on the issue of quasar accretion disc temperature profiles
will require additional and improved observations and modelling.  Measurements
of the wavelength dependence of half-light radii in lensed quasars could be
expanded to larger samples
and extended to shorter wavelengths using the Hubble Space Telescope.
Improved $M_{BH}$ estimates for gravitationally lensed quasars would yield
better constraints on models providing acceptable fits to half-light radii,
as would multiwavelength SED measurements 
(such as photometrically calibrated flux measurements of lens images
at observed optical and infrared wavelengths).
One worthwhile improvement to the modelling would be to self-consistently use
the luminosity profiles which have here been found to provide good fits to the
half-light radii as input to simulations of the microlensing flux deviations.
Currently the simulations use elliptical Gaussians, whereas at high
temperatures the best-fit profiles found here
will resemble elliptical rings on the sky.

Further exploration of model temperature profiles consistent with improved
observations could then follow.
Models with \fedd$>$1 and super-Eddington luminosities
\citep[e.g.,][]{1988ApJ...332..646A} may be worth considering, as the
best-fit models we find here tend to cluster at our imposed limit of \fedd=1.
Also, each of our magnetically restrained models has an angular velocity which
is a fixed fraction $f$ of the Keplerian value, but a more general extension
of the models of \cite{1997MNRAS.288...63O} would involve a power-law
dependence of the angular velocity on radius ($f\equiv (R/R_b)^x$), arising
from an assumed radial power-law dependence of the magnetic field strength.

Eventually, such parametric modelling should be superceded by the use of 
temperature profiles from theoretical simulations of accretion discs.
If the results of B11 are borne out by further observations,
a strong test of such simulations will be how naturally they can
generate the temperature profiles allowed by observations.

Alternatively, if future observations rule out the possibility 
that quasar accretion discs have flat temperature profiles, 
inhomogeneous accretion discs \citep{da11} offer a potential solution
to the discrepancy between microlensing and flux sizes.
Inhomogeneous discs simultaneously increase the radius within which 
emission at a given wavelength occurs (increasing the microlensing size)
and decrease the area within that radius which emits at that wavelength
(increasing the flux size).

\section*{Acknowledgements}

We thank C. Kochanek for valuable comments.
PBH thanks NSERC for research support for LSC and EW.
ESN thanks NSERC for an Undergraduate Student Research Award.
CJN acknowledges support provided by NASA through the Einstein Fellowship Program, grant PF2-130098.

\bibliographystyle{mn}

\appendix
\section{Forming disc model constraints} \label{fdcon}

The energy per second emitted from the disc in its forming region ($R>R_b$)
must satisfy $L\leq\frac{1}{2}G\dot{M}M_{BH}/R_b$ (see \S \ref{fd}).

Considering both sides of the disc,
\begin{equation}
 2\int_{R_b}^\infty \sigma_{sb} [T_{\rm surf}(R)]^4 dA \leq \frac{GM_{BH}\dot{M}}{2R_b} \end{equation}
Writing $T_{\rm surf}(R)=T_{hi} (R/R_b)^{-\beta}$ at $R\geq R_b$,
with $\beta>\frac{1}{2}$, we have 
\begin{equation}
 2\int_{R_b}^\infty \sigma_{sb} T_{hi}^4(R/R_b)^{-4\beta} 2\pi R dR \leq \frac{GM_{BH}\dot{M}}{2R_b} \end{equation}
\begin{equation}
 4\pi R_b^2 \sigma_{sb} T_{hi}^4 \int_{R_b}^\infty (R/R_b)^{-4\beta} R/R_b d(R/R_b) \leq \frac{GM_{BH}\dot{M}}{2R_b} \end{equation}
Making the substitution $x=R/R_b$,
\begin{equation}
 4\pi R_b^2 \sigma_{sb} T_{hi}^4 \int_{1}^\infty x^{-4\beta+1} dx \leq \frac{GM_{BH}\dot{M}}{2R_b} \end{equation}
\begin{equation}
 4\pi R_b^2 \sigma_{sb} T_{hi}^4 \frac{1}{4\beta-2}\leq \frac{GM_{BH}\dot{M}}{2R_b} \end{equation}
\begin{equation}
 \sigma_{sb} T_{hi}^4 \leq \left[\beta-\frac{1}{2}\right] \frac{GM_{BH}\dot{M}}{2\pi R_b^3} \end{equation}

\section{Magnetically restrained disc model constraints} \label{mrcon}

In our magnetically restrained disc models, gas accreting through 
a Keplerian disc at large radii is presumed to reduce its orbital velocity 
to a fraction $f<1$ of its Keplerian value by the time it reaches radius $R_b$.
The disc thus has a fraction $(1-f^2)$ of the kinetic energy of the
accreting gas at $R_b$ available to emit as heat at $R>R_b$, in addition
to the normal disc emission ($\propto T_{\rm SS73}^4$) at $R>R_b$.

Consider a disc obeying
$T_{\rm surf}(R)=T_{hi} (R/R_b)^{-\beta}$ from $R_b\leq R\leq R_c$
and a normal SS73 $T_{\rm surf}(R)=T_{lo}(R/Rb)^{-3/4}$ profile at $R>R_c$,
where $T_{hi}(R_c/R_b)^{-\beta} \equiv T_{lo}(R_c/R_b)^{-3/4}$ so that
$T_{lo}/T_{hi}=(R_c/R_b)^{-\beta+3/4}$ and
$(R_c/R_b)=(T_{lo}/T_{hi})^{\frac{1}{-\beta+3/4}}$.

We have a limit on the excess energy emitted by the disc, which is equal to the
total energy emitted by the disc minus the energy that would have been emitted
by an SS73 disc between those radii.  Again considering both sides of the disc, 
\begin{eqnarray}
 2\int_{R_b}^{R_c} \sigma_{sb} T_{hi}^4(R/R_b)^{-4\beta} dA
 - 2\int_{R_b}^{R_c} \sigma_{sb} T_{lo}^4(R/R_b)^{-3} dA \nonumber\\
\leq (1-f^2)\frac{GM_{BH}\dot{M}}{2R_b} \end{eqnarray}
Expanding $dA$ and making the substitution $x=R/R_b$,
\begin{eqnarray}
 4\pi R_b^2 \sigma_{sb}T_{hi}^4 \left[ \int_{1}^{R_c/R_b} x^{-4\beta+1} dx 
- \int_1^{R_c/R_b} \frac{T_{lo}^4}{T_{hi}^4} x^{-2} dx \right] \nonumber\\
\leq (1-f^2)\frac{GM_{BH}\dot{M}}{2R_b} \end{eqnarray}
which evaluates to
\begin{eqnarray}
 4\pi R_b^2 \sigma_{sb} T_{hi}^4 
\left[ \frac{x^{-4\beta+2}}{-4\beta+2}\bigg|_1^{R_c/R_b}
+\left(\frac{T_{lo}}{T_{hi}}\right)^4 \frac{x^{-3}}{3}\bigg|_1^{R_c/R_b}
\right] \nonumber\\
\leq (1-f^2)\frac{GM_{BH}\dot{M}}{2R_b} \end{eqnarray}

Defining $t_{lh}\equiv T_{lo}/T_{hi}$, we have:
\begin{eqnarray}
4\pi R_b^2 \sigma_{sb} T_{hi}^4 
\left[ 
\frac{1}{-4\beta+2}
\left(
t_{lh}^{\frac{4\beta-2}{\beta-3/4}}
-1\right) +\frac{1}{3} 
\left(
t_{lh}^{\frac{4\beta}{\beta-3/4}} 
-t_{lh}^4 
\right)
\right]\nonumber\\
 \leq (1-f^2)\frac{GM_{BH}\dot{M}}{2R_b} 
\end{eqnarray}
Examining the above equation,
we see that for it to be satisfied in the limit $f\rightarrow 1$,
where no excess energy is available, 
then we must have $t_{lh}\rightarrow 1$, or $T_{hi}=T_{lo}$, as expected.

Thus, the constraint on our magnetically restrained disc models
is given by:
\begin{eqnarray} \label{eqmrcon}
\sigma_{sb} T_{hi}^4 \leq
(1-f^2)\frac{GM_{BH}\dot{M}}{8\pi R_b^3} 
\bigg/
\left[ 
\frac{1}{-4\beta+2} \left( t_{lh}^{\frac{4\beta-2}{\beta-3/4}} -1\right) 
+\frac{1}{3} \left( t_{lh}^{\frac{4\beta}{\beta-3/4}} -t_{lh}^4 \right) 
\right] 
\end{eqnarray}
where $t_{lh}\equiv T_{lo}/T_{hi}$.

\section{Accretion disc flux sizes} \label{sizes}

The {\em flux size} of a quasar accretion disc (M10) is found 
by using the quasar's observed flux to calculate its luminosity
and determining the size a standard SS73 disc would need to be
to generate that luminosity.

Given a lensed quasar's demagnified apparent magnitude $I$,
the flux size in the $I$ band ($\lambda_{I,obs}=0.814$\,\micron) 
for the accretion disc of that quasar,
located at angular size distance $D_{OS}/r_H$ 
(in units of the Hubble radius $r_H=c/H_0$) with inclination $i$
and with a temperature profile $T(R)\propto R^{-\beta}$, 
is given by:
\begin{eqnarray}
R_I=\frac{2.83\times10^{15}}{\sqrt{K_\lambda(\beta)\cos i}}\left(\frac{D_{OS}}{r_{H}}\right)\left(\frac{\lambda_{I,obs}}{\mu \rm m}\right)^{3/2}10^{-0.2(I-19)}h^{-1}\rm~cm
\end{eqnarray}
where $H_0=100h$~\kmsm\ and where
\begin{eqnarray}
K_\lambda(\beta)=\frac{1}{2.58}\int_{0}^{\infty} u ~du ~[\exp(u^\beta)-1]^{-1} 
\end{eqnarray}
is the integral over the disc surface brightness,
normalized to unity for $\beta=3/4$.

The $K_\lambda(\beta)$ integral is obtained as follows.
For observations at rest wavelength $\lambda$,
we can write $T(R)=T_\lambda(R/R_\lambda)^{-\beta}$, 
with $T_\lambda\equiv hc/k\lambda$ and 
\begin{eqnarray}
R_\lambda\equiv \frac{1}{\pi^2}\left(\frac{45}{16}\frac{\lambda^4R_g\dot{M}}{h_p}\right)^{1/3}
\end{eqnarray}
where $R_g=GM_{BH}/c^2$.
We then have
\begin{eqnarray} \label{eqfluxsize}
K_\lambda(\beta) \propto
\int_{0}^{\infty} R ~dR ~[\exp(hc/k\lambda T(R))-1]^{-1} \\
\propto \int_{0}^{\infty} R ~dR ~[\exp(R^\beta/R_\lambda^\beta)-1]^{-1} \\
\propto R_\lambda^2 \int_{0}^{\infty} u ~du ~[\exp(u^\beta)-1]^{-1}
\end{eqnarray} 
where we have defined $u\equiv R/R_\lambda$.

In the case of our magnetically restrained models (\S~\ref{mr}), 
where the disc temperature profile has three parts,
we can generalize $K_\lambda(\beta)$ to $\kappa_\lambda(\beta,f,T_{lo},T_{hi})$:
\begin{eqnarray} \label{newK}
&&\kappa_\lambda(\beta,f,T_{lo},T_{hi}) \propto \\
&&\left[ 
\int_{0}^{u_1} \frac{ R_\lambda^2~ u ~du }{ \exp(u^{3/4}/\sqrt{f})-1 }
+\int_{u_1}^{u_2} \frac{ R_\lambda^2~ u ~du }{ \exp(gu^{\beta})-1 }
+\int_{u_2}^{\infty} \frac{ R_\lambda^2~ u ~du }{ \exp(u^{3/4})-1 }
\right] 
\nonumber
\end{eqnarray}
where $u\equiv R/R_\lambda$ as before.
In the equation above, we define
$u_1\equiv R_b/R_\lambda$ and $u_2\equiv R_c/R_\lambda$,
with $R_b$ and $R_c$ as defined in \S~\ref{mr},
and we define $g \equiv (T_\lambda/T_{hi})(R_\lambda/R_b)^\beta$.
The constant of proportionality is the same for 
$K_\lambda(\beta)$ and $\kappa_\lambda(\beta,f,T_{lo},T_{hi})$,
so the ratio of those two quantities gives the ratio of the
flux sizes for an SS73 model disc and a magnetically arrested one.

\label{lastpage}
\end{document}